\documentclass[pra,twocolumn,superscriptaddress]{revtex4-2}
\usepackage{graphicx,multirow,dsfont}
\usepackage{amsmath,bm}
\usepackage{amssymb}
\usepackage{mathbbol}
\usepackage{amsfonts}
\usepackage{mathrsfs}
\usepackage{float}
\usepackage{xcolor}
\usepackage{hyperref}
\definecolor{darkblue}{rgb}{0,0.0.1,0.3}
\definecolor{darkred}{rgb}{0.6,0.1,0}
\hypersetup{colorlinks,breaklinks,
linkcolor=darkred,urlcolor=darkblue,
anchorcolor=darkred,citecolor=
darkred,pdfauthor=ShChAr, pdftitle=ShChAr.Teleportation}
\usepackage{tikz}
\usepackage{mathbbol}
\usepackage[normalem]{ulem}

\newcommand{\ie}{\textit{i}.\textit{e}.}

\begin{document}
\title{Continuous variable quantum teleportation using
photon subtracted and photon added two mode squeezed 
coherent state}
\author{Shikhar Arora}
\email{shikhar.quantum@gmail.com}
\affiliation{Department of Physical Sciences,
Indian
Institute of Science Education and
Research Mohali, Sector 81 SAS Nagar,
Punjab 140306 India.}
\author{Chandan Kumar}
\email{chandan.quantum@gmail.com}
\affiliation{Department of Physical Sciences,
Indian
Institute of Science Education and
Research Mohali, Sector 81 SAS Nagar,
Punjab 140306 India.}
\author{Arvind}
\email{arvind@iisermohali.ac.in}
\affiliation{Department of Physical Sciences,
Indian
Institute of Science Education  and
Research Mohali, Sector 81 SAS Nagar,
Punjab 140306 India.}
\begin{abstract}
We consider non-Gaussian states generated by photon
subtraction (PS) and photon addition (PA) on two-mode
squeezed coherent (TMSC) states, as resource states for
continuous variable (CV) quantum teleportation (QT). To this
end, we derive the Wigner characteristic function for the
family of photon subtracted and photon added TMSC states,
which is then utilized to calculate the fidelity of
teleporting a single mode coherent state and a squeezed vacuum
state.  The analysis shows that while symmetric PS enhances
the fidelity of QT in an extensive range of squeezing,
asymmetric  PS enhances the performance marginally and only in
the low squeezing regime.  The addition operations on the
other hand are less useful, symmetric three-PA leads to a
marginal improvement while the other addition operations are
useless.  We have considered the actual experimental setup
for PS and PA operations and computed their success
probabilities which should be kept in mind while advocating
the use of these operations.  We could compute the fidelity
of QT for a broad range of states because we analytically
derived the Wigner characteristic function for these family
of states which we think will be useful for various other
applications of these families of states.
\end{abstract}
\maketitle
\section{Introduction}
The non-classical subset of,
two mode Gaussian states, in particular, two mode squeezed
vacuum (TMSV) state, have played a crucial role in the
development of various continuous variable (CV) quantum
information processing (QIP) protocols, for instance,
quantum teleportation~\cite{bk-1998}, entanglement
swapping~\cite{swapping-1999}, and quantum key
distribution~\cite{Laudenbach-review}. However, in recent
times, much attention is being paid to non-Gaussian states,
which are generated using non-Gaussian operations such as
photon subtraction (PS) and  photon addition
(PA) on initial Gaussian states~\cite{Walschaers-prx-2021}.
These non-Gaussian operations can enhance the
nonclassicality~\cite{Agarwal-pra-1991} and entanglement
content of the state that they act upon
~\cite{Kitagawa-pra-2006,Ourjoumtsev-prl-2007,
Takahashi-nature-2010,Zhang-pra-2010,pspra2012}.
Non-Gaussian operations have already been considered for
performance enhancement in quantum
illumination~\cite{ill2008,ill2013,illumination14,illumination18,rivu},
quantum teleportation
(QT)~\cite{tel2000,Akira-pra-2006,Anno-2007,tel2009,wang2015,catalysis15,catalysis17,ayan,tele-arxiv,better}
and quantum
metrology~\cite{gerryc-pra-2012,josab-2012,braun-pra-2014,josab-2016,pra-catalysis-2021,crs-ngtmsv-met,ngsvs,metro-thermal}.
Furthermore, non-Gaussian operations have been utilized in
loophole-free tests of Bells
inequality~\cite{loophole-prl-2004,Grangier-prl-2004} and
in quantum computing~\cite{Bartlett-pra-2002}.

While dealing with Gaussian states and Gaussian operations
is theoretically simple, calculations involving non-Gaussian
states and non-Gaussian operations can be complicated. This
poses a formidable challenge to analyzing non-Gaussian
operations, even on simple Gaussian states.  The PS and PA
operations on TMSV states have been extensively studied in
quantum teleportation, whereas PS and PA operations on more
general class of Gaussian states such as two mode squeezed
coherent (TMSC) states are unexplored. This can be
attributed mainly to challenges involved in  the
mathematical description of these general non-Gaussian
states and in calculating the quantities of interest such as
QT fidelity.

It should be noted that PS and PA are probabilistic
processes, and therefore, the success probability must be
taken into account while analyzing any quantity of interest.
However, most previous research has considered ideal PS and
PA operations via the annihilation and creation operators
and ignored the success probability entirely. To account for
the success probability, one needs to consider the practical
scheme for non-Gaussian operations, which results in
additional parameters.  For instance, transmissivity
parameters are included while considering the PS operation
modeled using a beam splitter.  This additional parameter
further increases the challenge of working out an analytical
solution.

This work considers non-Gaussian states generated by
realistic PS and PA operations on TMSC state as resource
states for CV QT. TMSC state can be easily generated in a
lab by feeding coherent state to  a parametric down
converter~\cite{tmsc-pra-2015}. TMSC states have already
been studied in detail~\cite{caves-pra-91,simon-pra-94} and
recently, theoretical analysis was undertaken in the context
of quantum metrology~\cite{tmsc-pra-2015}.  Here we endeavor
to explore whether the state generated by
 PS and PA
operations on TMSC state when used as resource states
can yield higher fidelity of QT.
To this end, we first derive the Wigner characteristic
functions of the PSTMSC and PATMSC states in terms of
two-variable Hermite polynomials. These functions have the
transmissivity of the two beam splitters as parameters, and
by suitably choosing the transmissivity, we can perform
asymmetric and symmetric PS and PA operations. We then
derive the fidelity of QT for an input coherent state and
squeezed vacuum state. Since we have considered practical
models for PS and PA operations, we take the success
probability of the respective non-Gaussian operations into
account while analyzing the fidelity of QT.

The results show that symmetric PS operations enhance the
fidelity of QT.  However,  for the asymmetric PS
operations,  the enhancement is seen only
in the  low squeezing range. We note that
the success probability for symmetric $n$-PS is less when
compared to asymmetric $n$-PS. Among the PA operations,
symmetric 3-PA, whose success probability is very low,
slightly improves the fidelity of QT for an input squeezed
vacuum state. For all other cases, the PA operation is not
useful. We were able to carry out fidelity calculations
because we managed to compute the  the Wigner characteristic
function for the PSTMSC and PATMSC states which we think
would be valuable in characterizing these states and to
analyse their use in various QIP tasks.

The paper is organized as follows. In Sec.~\ref{sec:char},
we derive the Wigner characteristic functions of the PSTMSC
and PATMSC states. In Sec~\ref{sec:qt}, we derive the
fidelity of teleportation for an input coherent state and
squeezed vacuum state and analyze them. Finally, in
Sec.~\ref{sec:conc}, we provide some concluding remarks and
discuss future directions. In Appendix~\ref{intro}, we
provide a brief overview of CV systems and their phase space
representation.
\section{Wigner characteristic function of PSTMSC and PATMSC
states}
\label{sec:char}
We first derive the Wigner characteristic function of the
PSTMSC state.
\begin{figure}[h!] 
\includegraphics[scale=1]{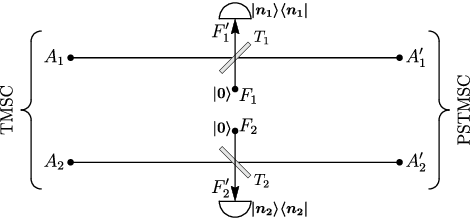}
\caption{Scheme for the preparation of PSTMSC state. Modes
$A_1$ and $A_2$ of TMSC state are mixed with modes $F_1$ and
$F_2$, initialized to vacuum states, using beam splitter of
transmissivity $T_1$ and $T_2$, respectively. Photon number
resolving detectors (PNRD) given by the POVM
$\{|n_1\rangle\langle n_1|,\mathbb{1}-|n_1\rangle\langle
n_1|\}$ and $\{|n_2\rangle\langle
n_2|,\mathbb{1}-|n_2\rangle\langle n_2|\}$ are applied to
modes $F_1^{'}$ and $F_2^{'}$ respectively. Simultaneous
click of the POVM elements $|n_1\rangle\langle n_1|$ and
$|n_2\rangle\langle n_2|$ represents successful subtraction
of $n_1$ and $n_2$ photons from the modes $A_1$ and $A_2$ of
the TMSC state, respectively.}
\label{figsub}
\end{figure}
The schematic for the generation of the PSTMSC state is
shown in Fig.~\ref{figsub}. We start with an uncorrelated
two mode coherent state given by
\begin{equation}
|\psi\rangle_{A_1 A_2}=D_{A_1}(d,d)  D_{ A_2}(d,d) |00\rangle,
\end{equation}
where $D_{i}(d,d)$ is the displacement operator defined in
Eq.~(\ref{dis}) of the Appendix~\ref{intro}. Since two-mode
coherent state is a Gaussian state, the Wigner
characteristic function can be written using
Eq.~(\ref{wigc}) as
\begin{equation}
\begin{aligned}
\chi (\Lambda)= & \exp \bigg[
-\frac{\tau_1^2+\sigma_1^2+\tau_2^2+\sigma_2^2}{4}   \\
& + i (\sigma_1+\sigma_2)\, d  -i (\tau_1+\tau_2)\, d  \bigg].
\end{aligned}
\end{equation}
The TMSC state is generated by sending two-mode coherent
state through a non-linear optical down
converter~\cite{tmsc-pra-2015}:
\begin{equation}
|\Psi\rangle_{A_1 A_2}=\mathcal{U}(S_{A_1 A_2}(r))|\psi\rangle_{A_1 A_2},
\end{equation} 
where $\mathcal{U}(S_{A_1 A_2}(r))$ is the two mode squeezing operation 
defined in Eq.~(\ref{twomodesq}) of the Appendix~\ref{intro}. The 
Wigner characteristic function transforms as 
$\chi(\Lambda)\rightarrow \chi(S_{A_1 A_2}^{-1}(r)\Lambda)$, which 
evaluates to
\begin{equation}
\begin{aligned}
& \chi_{A_1 A_2}(\Lambda)= \exp \bigg[i
(\sigma_1+\sigma_2)\, d \, e^r-i (\tau_1+\tau_2)\, d \,
e^{-r}   \\
&-\frac{\tau_1^2+\sigma_1^2+\tau_2^2+\sigma_2^2}{4} \cosh(2
r)+\frac{\tau_1 \tau_2-\sigma_1 \sigma_2}{2}\sinh(2 r)
\bigg].
\end{aligned}
\end{equation}
We combine the modes $A_1$ and $A_2$ of the TMSC state with ancilla modes 
$F_1$ and $F_2$, initiated to vacuum states using beam-splitters of 
transmissivity $T_1$ and $T_2$ respectively. We represent the modes $A_1$ 
and $A_2$ by the quadrature operators $(\hat{q}_1,\hat{p}_1)^T$ and 
$(\hat{q}_2,\hat{p}_2)^T$, and the  ancilla modes $F_1$ and $F_2$ by the 
quadrature operators $(\hat{q}_3,\hat{p}_3)^T$ and 
$(\hat{q}_4,\hat{p}_4)^T$, respectively. The characteristic function of 
the four mode system before the beam splitter operations can be written as
\begin{equation}
\chi_{F_1 A_1 A_2 F_2}(\Lambda) =  \chi_{A_1 A_2}(\Lambda)
\chi_{|0\rangle}(\Lambda_3)  \chi_{|0\rangle}(\Lambda_4),
\end{equation}
where $\chi_{|0\rangle}(\Lambda_i)$ $(i=3,4)$ is the Wigner characteristic 
function of the vacuum state. The four modes get entangled as a result of 
mixing by the two beam splitters 
$ B(T_1,T_2)=B_{A_1 F_1}(T_1) \oplus B_{A_2 F_2} (T_2)$, where $B_{ij}(T)$ 
is beam splitter operation defined in Eq.~(\ref{beamsplitter}) of the 
Appendix~\ref{intro}. The transformed characteristic function can be 
calculated as
\begin{equation}
\chi_{F_1' A_1' A_2' F_2'}(\Lambda)  =\chi_{F_1 A_1 A_2 F_2}( B(T_1,T_2)^{-1}\Lambda).
\end{equation}
The modes $F_1^{'}$ and $F_2^{'}$ are measured with photon number resolving 
detectors (PNRD) represented by the positive-operator-valued measure (POVM) 
$\{\Pi_{n_1}=|n_1\rangle\langle n_1|,\mathbb{1}-\Pi_{n_1}\}$ and 
$\{\Pi_{n_2}=|n_2\rangle\langle n_2|,\mathbb{1}-\Pi_{n_2}\}$ respectively. 
When the POVM elements $\Pi_{n_1}$ and $\Pi_{n_2}$ click simultaneously, PS 
operation on both the modes is considered to be successful. In this paper, 
we consider symmetric and asymmetric PS operation. Symmetric PS corresponds 
to equal number of photons being detected in both the modes, \ie, 
$n_1=n_2=n$. The resultant state is referred to as Sym $n$-PSTMSC state. 
Asymmetric PS corresponds to PS on one of the modes (mode $A_2$ in this 
paper). The resultant state is referred to as Asym $n$-PSTMSC state. The 
results for Asym $n$-PSTMSC state can be obtained from the general 
expression by setting $n_1=0$,  $n_2=n$, and $T_1 = 1$. Post-measurement, 
the unnormalized Wigner characteristic function can be written as
\begin{equation}\label{detect}
\begin{aligned}
\widetilde{\chi}^{\text{PS}}_{A_1' A_2'}=& \frac{1}{(2
\pi)^2} \int d^2 \Lambda_3  d^2 \Lambda_4 
\underbrace{\chi_{F_1' A_1' A_2' F_2'}(\Lambda)}_{\text{Four mode entangled state}}\\
&\times 
\underbrace{\chi_{|n_1\rangle
}(\Lambda_3)}_{\text{Projection on }|n_1\rangle \langle
n_1|} 
\underbrace{\chi_{|n_2\rangle
}(\Lambda_4)}_{\text{Projection on }|n_2\rangle \langle
n_2|}, \\
\end{aligned}
\end{equation}
where $\chi_{|n_i\rangle }(\Lambda_i)$ ($i=3,4$) is the Wigner 
characteristic function of the Fock state $|n_i\rangle$ and can be 
written in terms of Laguerre polynomial as given in Eq.~(\ref{charfock1}) 
of the Appendix~\ref{intro}. Integration of Eq.~(\ref{detect}) yields
\begin{equation}\label{eqchar}
\begin{aligned}
\widetilde{\chi}^{\text{PS}}_{A_1' A_2'}&=a_0 \exp
\left(\bm{\Lambda}^T M_1 \bm{\Lambda}+\bm{\Lambda}^T
M_2\right)\\
&\times \bm{\widehat{F}_1} \exp  \big[-a_1 u_1 v_1+a_2
u_1+a_3 v_1 -a_4 u_2 v_2  \\
& +a_5 u_2+a_6 v_2+ a_7  (u_1 u_2+v_1 v_2 ) \big] ,\\
\end{aligned}
\end{equation}
where the coefficients $a_i$ are provided in Eq.~(\ref{ai}) of the 
Appendix~\ref{appps}, the column vector $\bm{\Lambda}$ is defined as 
$(\tau_1,\sigma_1,\tau_2,\sigma_2)^T$, and the matrices $M_1$ and $M_2$ 
are given in Eqs.~(\ref{ai3}) and~(\ref{ai33}) of the Appendix~\ref{appps}. 
Further, the differential operator $\bm{\widehat{F}_1} $ is defined as
\begin{equation}
\begin{aligned}
\bm{\widehat{F}_1} = \frac{2^{-(n_1+n_2)}}{n_1!n_2!}
\frac{\partial^{n_1}}{\partial\,u_1^{n_1}}
\frac{\partial^{n_1}}{\partial\,v_1^{n_1}} 
\frac{\partial^{n_2}}{\partial\,u_2^{n_2}}
\frac{\partial^{n_2}}{\partial\,v_2^{n_2}} \{ \bullet
\}_{\substack{u_1= v_1=0 \\ u_2= v_2=0}}.\\
\end{aligned}
\end{equation}
Equation~(\ref{eqchar}) can further be expressed in terms of two-variable 
Hermite polynomials $H_{m,n}(x,y)$, which has been explicitly demonstrated 
in Appendix~\ref{appps}:
\begin{equation}\label{eq4}
\begin{aligned}
&\widetilde{\chi}^{\text{PS}}_{A_1' A_2'}= a_0
\frac{2^{-(n_1+n_2)}}{n_1!n_2!} \exp \left(\bm{\Lambda}^T
M_1 \bm{\Lambda}+\bm{\Lambda}^T M_2\right)\\
&\times \sum_{i,j=0}^{\text{min}(n_1,n_2)}   \frac{a_1^{n_1}}{\sqrt{a_1}^{i+j}} 
\frac{a_4^{n_2}}{\sqrt{a_4}^{i+j}} \frac{a_7^{i+j}}{i!\,j!}
P^{n_1}_i P^{n_1}_j P^{n_2}_i P^{n_2}_j \\
&\times
H_{n_1-i,n_1-j}\left[\frac{a_2}{\sqrt{a_1}},\frac{a_3}{\sqrt{a_1}}\right]
H_{n_2-i,n_2-j}\left[\frac{a_5}{\sqrt{a_4}},\frac{a_6}{\sqrt{a_4}}\right],
\\
\end{aligned}
\end{equation}
where $P^n_r=n!/(n-r)!$ is $r$-permutation of $n$. We note that the 
afore-derived Wigner characteristic function is unnormalized. The 
normalization corresponds to the probability of simultaneous  click of 
both the PNRD detectors and can be computed as follows:
\begin{equation}\label{probps}
\begin{aligned}
P^{\text{PS}}&=\widetilde{\chi}^{\text{PS}}_{A_1'
A_2'}\bigg|_{\substack{\tau_1= \sigma_1=0\\ \tau_2=
\sigma_2=0}}
=a_0 \bm{\widehat{F}_1} \exp  \big[-a_1 u_1 v_1+b_2 u_1\\
&+b_3 v_1  -a_4 u_2 v_2   +b_5 u_2+b_6 v_2+a_7 \left(u_1
u_2+v_1 v_2\right) \big], \\
\end{aligned}
\end{equation}
where the coefficients $b_i$ are provided in Eq.~(\ref{ai2}) of the 
Appendix~\ref{appps}. Equation~(\ref{probps}) can also be expressed in 
terms of two-variable Hermite polynomials as
\begin{equation}\label{probps:her}
\begin{aligned}
&P^{\text{PS}}=
a_0 \frac{2^{-(n_1+n_2)}}{n_1!n_2!}
\sum_{i,j=0}^{\text{min}(n_1,n_2)}
\frac{a_1^{n_1}}{\sqrt{a_1}^{i+j}} 
\frac{a_4^{n_2}}{\sqrt{a_4}^{i+j}} \frac{a_7^{i+j}}{i!\,j!}  P^{n_1}_i \\
&\times P^{n_1}_j P^{n_2}_i P^{n_2}_j
H_{\substack{n_1-i,\\n_1-j}}\left[\frac{b_2}{\sqrt{a_1}},\frac{b_3}{\sqrt{a_1}}\right]
H_{\substack{n_2-i,\\n_2-j}}\left[\frac{b_5}{\sqrt{a_4}},
\frac{b_6}{\sqrt{a_4}}\right]. \\
\end{aligned}
\end{equation}

\begin{figure}[h!] 
\includegraphics[scale=1]{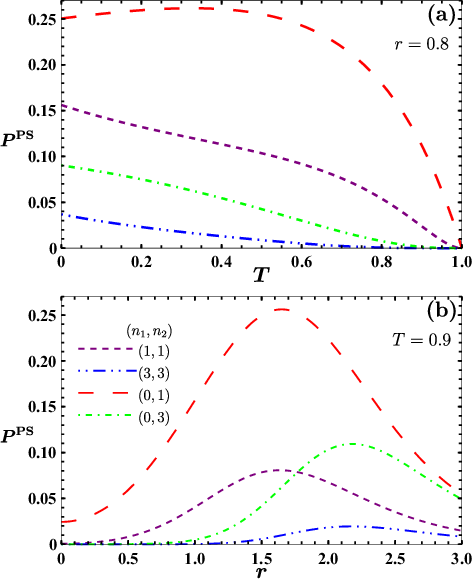}
\caption{Success probability $P^\text{PS}$ of PSTMSC states
as a function of (a)  transmissivity $T$ and (b)squeezing
parameter $r$.  The displacement has been taken as $d=0.5$.
$(n_1,n_2)$ represents the number of photons subtracted from
each mode.}
\label{prob_sub_r}
\end{figure}

We now analyze the success probability for various PSTMSC
states as a function of transmissivity and squeezing
parameter  in Fig.~\ref{prob_sub_r}. For numerical analysis,
we have assumed $T_1=T_2=T$ throughout this paper.  We see
that the success probability decreases as transmissivity
increases and approaches zero at unit transmissivity.  The
results show an optimum squeezing at which the probability
is maximum. We notice that the Asym $n$-PSTMSC and Sym
$n$-PSTMSC states achieve the maximum around the same
squeezing; however, the magnitude of the success probability
is less for Sym $n$-PSTMSC state as compared to Asym
$n$-PSTMSC state. Furthermore, the success probability for
Asym $n$-PSTMSC and Sym $n$-PSTMSC states decreases as $n$
increases. We also noticed from our numerical analysis (not
shown) that the maximum shifts to lower squeezing as the
transmissivity decreases.

The normalized Wigner characteristic function
$\chi^{\text{PS}}_{A'_1 A'_2}$ 
of the non-Gaussian PSTMSC state is obtained as
\begin{equation}\label{normPS}
\chi^{\text{PS}}_{A'_1
A'_2}(\tau_1,\sigma_1,\tau_2,\sigma_2)
={\left(P^{\text{PS}}\right)}^{-1}\widetilde{\chi}^{\text{PS}}_{A_1'
A_2'}(\tau_1,\sigma_1,\tau_2,\sigma_2).
\end{equation}
Several special cases can be obtained from the above
expression. The Wigner characteristic function of the ideal
PSTMSC state $\hat{a}_1^{n_1} \hat{a}_2^{n_2}
|\text{TMSC}\rangle$ can be obtained by setting $T_1=T_2=1$
in Eq.~(\ref{normPS}). Further, setting $d=0$, and
$T_1=T_2=1$ in Eq.~(\ref{normPS}) yields the Wigner
characteristic function of the ideal PSTMSV state
$\hat{a}_1^{n_1} \hat{a}_2^{n_2} |\text{TMSV}\rangle$. Now
we derive the Wigner characteristic function of PATMSC
state.

\begin{figure}[h!] 
\includegraphics[scale=1]{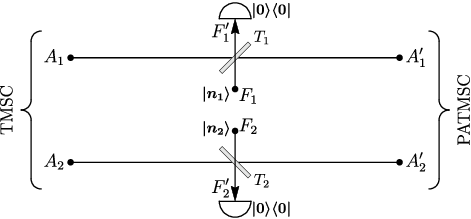}
\caption{Scheme for the preparation of PATMSC state. Modes
$A_1$ and $A_2$ of TMSC state is mixed with modes $F_1$ and
$F_2$, initialized to Fock states $|n_1\rangle$ and
$|n_2\rangle$, using beam splitter of transmissivity $T_1$
and $T_2$, respectively.  On-off detectors given by the POVM
$\{|0\rangle\langle 0|,\mathbb{1}-|0\rangle\langle 0|\}$ and
$\{|0\rangle\langle 0|,\mathbb{1}-|0\rangle\langle 0|\}$ are
applied to modes  $F_1^{'}$ and $F_2^{'}$ respectively.
Simultaneous click of the `on' elements of both the
detectors represents successful addition of $n_1$ and $n_2$
photons to the TMSC state.}
\label{figadd}
\end{figure}
The schematic for the generation of PATMSC state is
portrayed in Fig.~\ref{figadd}. We mix the modes $A_1$ and
$A_2$ of the TMSC state, with ancilla modes $F_1$ and $F_2$,
initiated to Fock states $|n_1\rangle$ and $|n_2\rangle$,
using beam-splitters of transmissivity $T_1$ and $T_2$
respectively. The Wigner characteristic function of the four
mode system post the beam splitter operations can be written
as
\begin{equation}
\chi_{F_1 A_1 A_2 F_2}(\Lambda) =  \chi_{A_1 A_2}(\Lambda_1,
\Lambda_2) \chi_{|n_1\rangle}(\Lambda_3)
\chi_{|n_2\rangle}(\Lambda_4).
\end{equation}
The two beam splitters entangle the four modes. The Wigner characteristic 
function after the beam-splitter operations is given by
\begin{equation}
\chi_{F_1' A_1' A_2' F_2'}(\Lambda)  =\chi_{F_1 A_1 A_2
F_2}( B(T_1,T_2)^{-1}\Lambda).
\end{equation}
The modes $F_1^{'}$ and $F_2^{'}$ are measured with on-off
detectors represented by the POVM
$\{\Pi_{0}=|0\rangle\langle 0|,\mathbb{1}-\Pi_{0}\}$. When
both the POVM elements  $\Pi_{0}$ click simultaneously,
$n_1$ and $n_2$ photons are considered to be added to the
TMSC state. Analogous to the PS case, we consider symmetric
PA for $n_1=n_2=n$ and asymmetric PA (on mode $A_2$) for
$n_1=0$, $n_2=n$ and $T_1 =1$. The corresponding states are
called Sym $n$-PATMSC and Asym $n$-PATMSC  states. The
unnormalized Wigner characteristic function for the PATMSC
state can be written as
\begin{equation}\label{detectPA}
\begin{aligned}
\widetilde{\chi}^{\text{PA}}_{A_1' A_2'}=&\frac{1}{(2
\pi)^2} \int d^2 \Lambda_3  d^2 \Lambda_4
\underbrace{\chi_{F_1' A_1' A_2' F_2'}(\Lambda)}_{\text{Four
mode entangled state}}\\
&\times 
\underbrace{\chi_{|0\rangle }(\Lambda_3)}_{\text{Projection
on }|0\rangle \langle 0|} \underbrace{\chi_{|0\rangle
}(\Lambda_4)}_{\text{Projection on }|0\rangle \langle 0|}.
\\
\end{aligned}
\end{equation}
On integrating Eq.~(\ref{detectPA}), we obtain  
\begin{equation}\label{unPA}
\begin{aligned}
\widetilde{\chi}^{\text{PA}}_{A_1' A_2'}&=a_0 \exp
\left(\bm{\Lambda}^T M_1 \bm{\Lambda}+\bm{\Lambda}^T
M_2\right)\\
&\times \bm{\widehat{F}_1} \exp  \big[-c_1 u_1 v_1+c_2
u_1+c_3 v_1-c_4 u_2 v_2\\
&+c_5 u_2+c_6 v_2+c_7 \left(u_1 u_2+v_1 v_2\right)\big],\\
\end{aligned}
\end{equation}
where the coefficients $c_i$ are provided in Eq.~(\ref{ai4})
of the Appendix~\ref{appps}. The afore-derived expression
can also be written in terms of two variable Hermite
polynomial as
\begin{equation}\label{eqpac}
\begin{aligned}
&\widetilde{\chi}^{\text{PA}}_{A_1' A_2'}= a_0
\frac{2^{-(n_1+n_2)}}{n_1!n_2!} \exp \left(\bm{\Lambda}^T
M_1 \bm{\Lambda}+\bm{\Lambda}^T M_2\right)\\ 
&\times \sum_{i,j=0}^{\text{min}(n_1,n_2)}
\frac{c_1^{n_1}}{\sqrt{c_1}^{i+j}} 
\frac{c_4^{n_2}}{\sqrt{c_4}^{i+j}} \frac{c_7^{i+j}}{i!\,j!}
P^{n_1}_i P^{n_1}_j P^{n_2}_i P^{n_2}_j \\
&\times
H_{n_1-i,n_1-j}\left[\frac{c_2}{\sqrt{c_1}},\frac{c_3}{\sqrt{c_1}}\right]
H_{n_2-i,n_2-j}\left[\frac{c_5}{\sqrt{c_4}},\frac{c_6}{\sqrt{c_4}}\right].
\\
\end{aligned}
\end{equation}
The probability of $n_1$ and $n_2$ photon addition on the
TMSC state can be evaluated from the above
equation~(\ref{eqpac}) as follows:
\begin{equation}\label{probpa}
\begin{aligned}
P^{\text{PA}}&=\widetilde{\chi}^{\text{PA}}_{A_1'
A_2'}\bigg|_{\substack{\tau_1= \sigma_1=0 \\ \tau_2=
\sigma_2=0}}=a_0 \bm{\widehat{F}_1} \exp  \big[-c_1 u_1
v_1+d_2 u_1\\
&+d_3 v_1 -c_4 u_2 v_2+d_5 u_2+d_6 v_2+c_7 \left(u_1 u_2+v_1 v_2\right)\big],\\
\end{aligned}
\end{equation}
where the coefficients $d_i$ are given in Eq.~(\ref{ai5}) of the 
Appendix~\ref{appps}. Equation~(\ref{probpa}) can also be expressed in 
terms of two-variable Hermite polynomial:
\begin{equation} 
\begin{aligned}
&P^{\text{PA}}=
a_0 \frac{2^{-(n_1+n_2)}}{n_1!n_2!}
\sum_{i,j=0}^{\text{min}(n_1,n_2)}
\frac{c_1^{n_1}}{\sqrt{c_1}^{i+j}} 
\frac{c_4^{n_2}}{\sqrt{c_4}^{i+j}} \frac{c_7^{i+j}}{i!\,j!}  P^{n_1}_i \\
&\times P^{n_1}_j P^{n_2}_i P^{n_2}_j
H_{\substack{n_1-i,\\n_1-j}}\left[\frac{d_2}{\sqrt{c_1}},\frac{d_3}{\sqrt{c_1}}\right]
H_{\substack{n_2-i,\\n_2-j}}\left[\frac{d_5}{\sqrt{c_4}},\frac{d_6}{\sqrt{c_4}}\right].
&&\\
\end{aligned}
\end{equation}

\begin{figure}[h!] 
\includegraphics[scale=1]{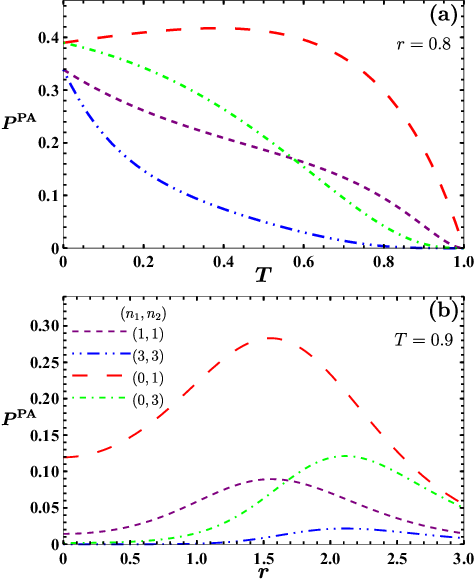}
\caption{Success probability $P^\text{PA}$ of PATMSC states
as a function of (a)  transmissivity $T$ and (b) squeezing
parameter $r$. The displacement has been taken as $d=0.5$.
$(n_1,n_2)$ represents the number of photons added to each
mode.}
\label{prob_add_r}
\end{figure}

Now we analyze the success probability for the PATMSC states
as a function of transmissivity and  squeezing parameter in
Fig.~\ref{prob_add_r}.  Again we see that the success
probability decreases as transmissivity increases and
approaches zero at unit transmissivity.  Further,  the Asym
$n$-PATMSC and Sym $n$-PATMSC states achieve a maximum
around the same squeezing, and the magnitude of success
probability for Sym $n$-PATMSC state is lower than the Asym
$n$-PATMSC state. Further, the success probability for Asym
$n$-PATMSC and Sym $n$-PATMSC states decrease as $n$
increases. We also observed that the maxima shifts to lower
squeezing as the transmissivity decreases.

The normalized  Wigner characteristic function
$\chi^{\text{PA}}_{A'_1 A'_2}$ of the non-Gaussian PATMSC
state is given by
\begin{equation}\label{normPA}
\chi^{\text{PA}}_{A'_1
A'_2}(\tau_1,\sigma_1,\tau_2,\sigma_2)
={\left(P^{\text{PA}}\right)}^{-1}\widetilde{\chi}^{\text{PA}}_{A_1'
A_2'}(\tau_1,\sigma_1,\tau_2,\sigma_2).
\end{equation}

We can obtain the Wigner characteristic function of several
special cases from the above expression. For instance, by
setting $T_1=T_2=1$ in Eq.~(\ref{normPA}), we get the Wigner
characteristic function of the ideal PATMSC state
$\hat{a}{_1^{\dagger }}^{m_1}\hat{a}{_2^{\dagger }}^{m_2}
|\text{TMSC}\rangle$.  On further setting $d=0 $, we obtain
the Wigner characteristic function of the ideal PATMSV state
$\hat{a}{_1^{\dagger }}^{m_1}\hat{a}{_2^{\dagger }}^{m_2}
|\text{TMSV}\rangle$.
\section{CV QT using PSTMSC and PATMSC states}\label{sec:qt}
After deriving the Wigner characteristic function of the
PSTMSC and PATMSC states, we derive the teleportation
fidelity to teleport a coherent and squeezed vacuum state.
\begin{figure}[h!] 
\includegraphics[scale=1]{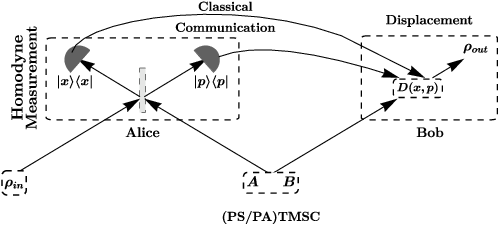}
\caption{Schematic of ideal Braunstein-Kimble   protocol for
the quantum teleportation of an unknown input quantum state.
The shared resource state can be either TMSC, PSTMSC or
PATMSC.}
\label{telescheme}
\end{figure}

Here we consider the ideal Braunstein-Kimble (BK) protocol
for the QT of an unknown input quantum state between two
distant physical systems~\cite{bk-1998}. The schematic is
shown in Fig.~\ref{telescheme}.  In the protocol, Alice and
Bob share an entangled resource state. The density operator
and the characteristic function of the entangled resource
state is represented by  $\rho_{A_1' A_2'}$ and $\chi_{A_1'
A_2'}(\Lambda_1, \Lambda_2)$, respectively. Alice has the
single mode input state that has to be teleported. We denote
the density operator and the characteristic function of the
unknown input state by $\rho_{\text{in}}$ and
$\chi_{\text{in}}(\Lambda_{\text{in}})$, respectively. Using
a balanced beam splitter, Alice mixes her mode and the
single mode input state. Alice then performs homodyne
measurements of $\hat{x}$ and $\hat{p}$-quadratures on the
two output modes of the beam splitter and classically
communicates the result to Bob. Bob performs a displacement
operation $D(x,p)$ on his mode, and consequently, the mode
$A_2'$ with Bob is transformed to mode `out'. The state
corresponding to the mode `out' is the teleported state. In
the characteristic function formalism, we can write the
teleported state as a product of the input state and the
entangled resource state~\cite{Marian-pra-2006}:
\begin{equation}\label{teleported}
\chi_{\text{out}}(\tau_2,\sigma_2) =
\chi_{\text{in}}(\tau_2,\sigma_2) \chi_{A_1'
A_2'}(\tau_2,-\sigma_2,\tau_2,\sigma_2).
\end{equation}

To assess the success of the QT protocol, we define fidelity
of QT as the overlap between the single mode  input state
$\rho_{\text{in}}$ and the output state $\rho_{\text{out}}$
as $F =\text{Tr} [\rho_{\text{in}}\rho_{\text{out}}]$. The
fidelity of QT in characteristic function can be written
as~\cite{Welsch-pra-2001}
\begin{equation}\label{fidex1}
F =\frac{1}{2 \pi} \int d^2 \Lambda_2  \chi_{\text{in}}(\Lambda_2)
\chi_{\text{out}}(-\Lambda_2).
\end{equation}
The maximum fidelity of teleporting a coherent state without
using a shared entangled state is
$1/2$~\cite{Braunstein-jmo-2000,Braunstein-pra-2001};
therefore, fidelity greater than $1/2$ signifies a success
for CV QT. We note that perfect teleportation, \ie,
teleportation with unit fidelity, can only occur with an
infinitely entangled resource state.

\subsection{Teleportation of input coherent state} 
Now we evaluate the fidelity of teleportation for an input
coherent state using entangled PSTMSC state~(\ref{normPS})
as a resource. The Wigner characteristic function for the
coherent state is provided in Eq.~(\ref{chi_coh}) of
Appendix~\ref{intro}. The expression for fidelity can be
evaluated using Eq.~(\ref{fidex1}), which turns out to be 
\begin{equation}\label{fidps}
\begin{aligned}
&F^\text{PS}=\frac{e_0}{P^\text{PS}}
\frac{2^{-(n_1+n_2)}}{n_1!n_2!}
\sum_{i,j=0}^{\text{min}(n_1,n_2)}
\frac{e_1^{n_1}}{\sqrt{e_1}^{i+j}} 
\frac{e_4^{n_2}}{\sqrt{e_4}^{i+j}} \frac{e_7^{i+j}}{i!\,j!}  P^{n_1}_i\\
& \times P^{n_1}_j P^{n_2}_i P^{n_2}_j
H_{\substack{n_1-i,\\n_1-j}}\left[\frac{e_2}{\sqrt{e_1}},\frac{e_3}{\sqrt{e_1}}\right]
H_{\substack{n_2-i,\\n_2-j}}\left[\frac{e_5}{\sqrt{e_4}},\frac{e_6}{\sqrt{e_4}}\right],
\\
\end{aligned}
\end{equation}
where the coefficients $e_i$ are given in Eq.~(\ref{ai6}) of
the Appendix~\ref{appps}. The afore-derived expression
represents the fidelity of QT for a very general family of
non-Gaussian resource states. By taking proper limits, we
can obtain the fidelity of QT for various special cases as
follows: (i) Ideal PS on TMSC state by setting $T_1=T_2=1$.
(ii) Realistic PS on TMSV state by setting $d=0$.  (iii)
Ideal PS on TMSV state by setting $d=0$ and $T_1=T_2=1$.
Further, we can obtain fidelity of QT for asymmetric PS
operations (on mode $A_2$) by setting $n_1=0$ and $T_1=1$ in
the aforementioned cases.

The fidelity of QT using TMSC resource state can be obtained
by taking 
$n_1=n_2=0$ and the $T_1 =T_2=1$ in Eq.~(\ref{fidps}):
\begin{equation}\label{fidtmsc}
F^\text{TMSC}=\frac{1+\lambda}{2}\exp
\left[d^2\left(\lambda-1\right)\right], \quad \lambda =
\tanh{r}.
\end{equation}
Setting $d=0$, we obtain the fidelity corresponding to the
TMSV resource state: $(1+\lambda)/2$~\cite{bk-1998}.
Similarly, in the unit transmissivity limit with
$n_1=n_2=1$, the fidelity results matches with the resource
state $\hat{a}_1 \hat{a}_2  |\text{TMSV}\rangle$ considered
in Ref.~\cite{tel2009}.  We see that the fidelity is
independent of displacement of the input coherent state and
depends upon the displacement and squeezing of the TMSC
state. Similarly, the fidelity of QT for an input coherent
state using PATMSC resource state~(\ref{normPA}) can be
evaluated using Eq.~(\ref{fidex1}): 
\begin{equation}\label{fidpa}
\begin{aligned}
&F^\text{PA}=\frac{e_0}{P^\text{PA}}
\frac{2^{-(n_1+n_2)}}{n_1!n_2!}
\sum_{i,j=0}^{\text{min}(n_1,n_2)}
\frac{f_1^{n_1}}{\sqrt{f_1}^{i+j}} 
\frac{f_4^{n_2}}{\sqrt{f_4}^{i+j}} \frac{f_7^{i+j}}{i!\,j!}  P^{n_1}_i \\
& \times P^{n_1}_j P^{n_2}_i P^{n_2}_j
H_{\substack{n_1-i,\\n_1-j}}\left[\frac{f_2}{\sqrt{f_1}},\frac{f_3}{\sqrt{f_1}}\right]
H_{\substack{n_2-i,\\n_2-j}}\left[\frac{f_5}{\sqrt{f_4}},\frac{f_6}{\sqrt{f_4}}\right],
\\
\end{aligned}
\end{equation}
where the coefficients $f_i$ are given in Eq.~(\ref{ai7}) of the 
Appendix~\ref{appps}. 
The fidelity results matches with the resource state
$\hat{a}{_1^{\dagger }} \hat{a}{_2^{\dagger }}
|\text{TMSV}\rangle$ obtained in the unit transmissivity
limit with $n_1=n_2=1$~\cite{tel2009}.
\begin{figure}[h!]
\includegraphics[scale=1]{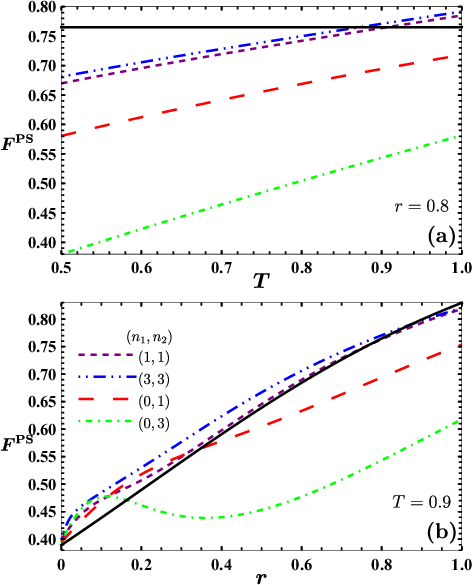}
\caption{Teleportation fidelity $F^\text{PS}$ for an input
coherent state using PSTMSC resource states as a function of
(a) transmissivity $T$ and (b) squeezing parameter $r$. The
displacement has been taken as $d=0.5$. The black solid
curve represents fidelity of teleportation for an input
coherent state using the TMSC resource state.}
\label{subfid}
\end{figure}

\begin{figure}[h!] 
\includegraphics[scale=1]{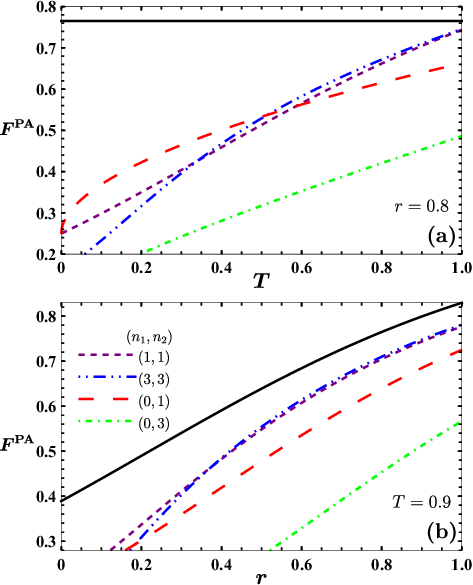}
\caption{ 
Teleportation fidelity $F^\text{PA}$ for an input coherent
state using PATMSC resource states as a function of (a)
transmissivity $T$ and (b) squeezing parameter $r$. The
displacement has been taken as $d=0.5$. The black solid
curve represents fidelity of teleportation for an input
coherent state using the TMSC resource state.}
\label{addfid}
\end{figure}
We numerically analyze the fidelity of QT for an input
coherent state as a function of transmissivity and squeezing
for the PSTMSC resource state and plot these results in
Fig.~\ref{subfid}.

From Fig.~\ref{subfid}(a), we observe that the fidelity of
QT increases as transmissivity increases; however, as we
noticed in Fig.~\ref{prob_sub_r}(b) that  the success
probability approaches zero  in the unit transmissivity
limit. Therefore, we need to work in the $T<1$ regime. For
symmetric operations, the fidelity of QT is above $1/2$ for
all values of transmissivity for the considered squeezing
parameter $r=0.8$. According to BK protocol, the excess
value beyond the classical bound $1/2$ is an indicator of
the success of CV QT. While Asym 1-PSTMSC state surpasses
the $1/2$ limit for all transmissivity values, Asym 3-PSTMSC
only surpasses the $1/2$ limit for $T \approx 0.8$. Further,
the teleportation fidelity for Sym 1-PSTMSC and Sym 3-PSTMSC
states outperforms the TMSC state after a threshold
transmissivity $T \approx 0.9$.

In Fig.~\ref{subfid}(b), we see that Sym 1-PSTMSC and Sym
3-PSTMSC states outperform the TMSC state in the approximate
interval $r \in (0,0.8)$.  Moreover, the teleportation
fidelity for these states surpasses the classical limit
$1/2$ for $r \approx 0.15$, which is a positive result. Asym
1-PSTMSC state outperforms Sym 1-PSTMSC when the fidelity of
teleportation is over the classical bound $1/2$. However,
Asym 3-PSTMSC state outperforms Sym 1-PSTMSC when the
fidelity of teleportation is below the classical bound
$1/2$. Therefore this scenario does not reflect the success
of the CV QT protocol.

The success probability for Sym 1-PSTMSC and Sym 3-PSTMSC
states are of the order $10^{-2}$ and $10^{-5}$ at $r=0.6$.
For Asym 1-PSTMSC state the probability is of the order
$10^{-2}$ at $r=0.2$.

We study the teleportation fidelity of an input coherent
state using the PATMSC resource state. Figure~\ref{addfid}
shows the results. We observe that the symmetric and
asymmetric photon addition only weakens the teleportation
fidelity in the sense that PATMSC resource states
underperform compared to the TMSC resource state. However,
the PATMSC states do surpass the classical bound $1/2$,
reflecting success for CV QT.

\subsection{Teleportation of input squeezed vacuum state} 

\begin{figure}[h!]
\includegraphics[scale=1]{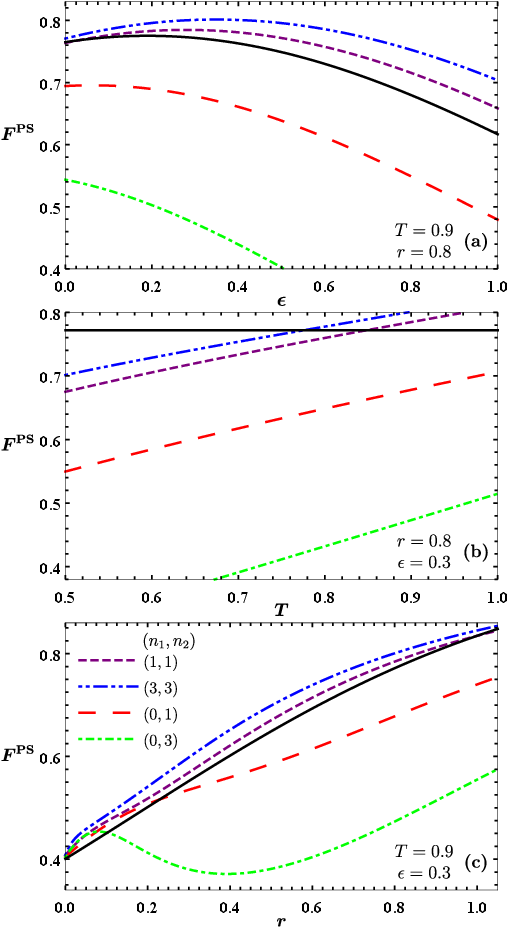}
\caption{(a) Teleportation fidelity $F^\text{PS}$ for input
squeezed vacuum state using PSTMSC resource states as a
function of the squeezing parameter $\epsilon$.  (b)
Teleportation fidelity $F^\text{PS}$ for an input squeezed
vacuum state using PSTMSC resource states as a function of
transmissivity $T$.  (c) Teleportation fidelity
$F^\text{PS}$ for an input squeezed vacuum state using
PSTMSC resource states as a function of squeezing parameter
$r$. The displacement has been taken as $d=0.5$. The black
solid curve represents fidelity of teleportation for an
input squeezed vacuum state using the TMSC resource state.}
\label{sqv_subtraction}
\end{figure}

We now consider the teleportation of input squeezed vacuum state with 
squeezing $\epsilon$ using entangled PSTMSC state as a resource state. 
The Wigner characteristic function of squeezed vacuum state can be 
obtained using Eq.~(\ref{chi_sqv}) of the Appendix~\ref{intro}. The 
expression for the fidelity of QT can be evaluated using 
Eq.~(\ref{fidex1}), which turns out to be 
\begin{equation}\label{sqvfidps}
F^\text{PS}=\frac{\bm{\widehat{F}_1} \exp \left(\bm{u}^T M_3
\bm{u}+\bm{u}^T M_4 + m_0\right)}{P^\text{PS}
\sqrt{b_0^2+d_0^2+2b_0d_0\cosh{(2\epsilon)}}},
\end{equation}
where $d_0 =(\cosh{r}-\sinh{r} \sqrt{T_1 T_2})^2  $ and the column 
vector $\bm{u}$ is defined as $(u_1,v_1,u_2,v_2)^T$. Further the 
expressions for $m_0$ and the matrices $M_3$ and $M_4$ are provided in 
Eqs.~(\ref{ai8}),~(\ref{ai9}), and~(\ref{ai10}) of the 
Appendix~\ref{appps}.

On taking $n_1=n_2=0$ and $T_1 =T_2 = 1$, we obtain fidelity of QT of an 
input squeezed vacuum state using TMSC resource state:
\begin{equation}\label{sqvfidtmsc}
\begin{aligned}
F^\text{TMSC}=&\left[\left(\frac{1+\tanh{(r+\epsilon)}}{2}\right)
\left(\frac{1+\tanh{(r-\epsilon)}}{2} \right)\right]^{1/2}\\
&\times \exp \left[d^2\left(\tanh{(r+\epsilon)}-1\right)\right],\\
\end{aligned}
\end{equation}
which depends on the squeezing $\epsilon$ of the input state and 
displacement and squeezing $r$ of the TMSC state. Further on taking 
$\epsilon=0$ in Eq.~(\ref{sqvfidtmsc}), we obtain Eq.~(\ref{fidtmsc}), 
the fidelity of QT of an input coherent state using TMSC resource state.

We now proceed to evaluate the fidelity of QT of an input squeezed vacuum 
state using PATMSC resource state, which evaluates to
\begin{equation}\label{sqvfidpa}
F^\text{PA}=\frac{\bm{\widehat{F}_1} \exp \left(\bm{u}^T M_5
\bm{u}+\bm{u}^T M_6 + m_0\right)}{P^\text{PA}
\sqrt{b_0^2+d_0^2+2b_0d_0\cosh{(2\epsilon)}}},
\end{equation}
where the explicit forms of the matrices $M_5$ and $M_6$ are provided in 
Eqs.~(\ref{ai11}), and~(\ref{ai12}) of the Appendix~\ref{appps}.

We analyze the fidelity of QT of an input squeezed vacuum
state using the PSTMSC resource state. We plot the
teleportation fidelity as a function of the squeezing
$\epsilon$ of the input squeezed vacuum state in
Fig.~\ref{sqv_subtraction}(a). The fidelity of QT for the
symmetric PSTMSC state and the TMSC state initially
increases, attains a maximum value, and then starts to
decrease as $\epsilon$ increases.

We plot the fidelity of QT as a function of transmissivity
in Fig.~\ref{sqv_subtraction}(b). We observe that the
performance of the symmetric PSTMSC state surpasses the TMSC
state after a threshold transmissivity $T \approx 0.8$. The
fidelity of QT as a function of squeezing parameter $r$ is
shown in Fig.~\ref{sqv_subtraction}(c). We observe that the
qualitative behavior of the fidelity for different PSTMSC
states is similar to that of Fig.~\ref{subfid}, the
teleportation fidelity for input coherent state.

\begin{figure}[h!] 
\includegraphics[scale=1]{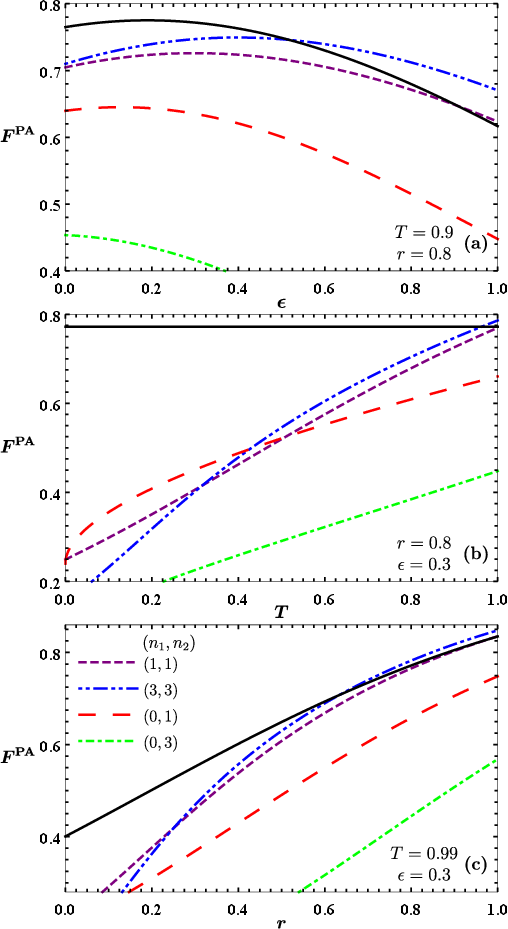}
\caption{(a) Teleportation fidelity $F^\text{PA}$ for input squeezed 
vacuum state using PATMSC resource states as a function of the 
squeezing parameter $\epsilon$. 
(b) Teleportation fidelity $F^\text{PA}$ for an input squeezed 
vacuum state using PATMSC resource states as a function of 
transmissivity $T$. 
(c) Teleportation fidelity $F^\text{PA}$ for an input squeezed 
vacuum state using PATMSC resource states as a function of 
squeezing parameter $r$. The displacement has been taken as 
$d=0.5$. The black solid curve represents fidelity of 
teleportation for an input squeezed vacuum state using the 
TMSC resource state.}
\label{sqv_addition}
\end{figure}

In Fig.~\ref{sqv_addition}, we analyze the teleportation
fidelity of input squeezed vacuum state using PATMSC
resource state. We observe from Fig.~\ref{sqv_addition}(b)
that the Sym 3-PATMSC state outperforms the TMSC state only
for high value of transmissivity ($T \approx 0.99$).
Further Fig.~\ref{sqv_addition}(c) shows that the Sym
3-PATMSC state outperforms the TMSC state for high squeezing
parameter. We note that the success probability for Sym
3-PATMSC is of the order $10^{-8}$ at $r=1$.

To conclude the analysis of fidelity of QT using PSTMSC and
PATMSC resource states, we find that the symmetric PS is the
most beneficial non-Gaussian operation. In
Table~\ref{table1}, we have provided the advantageous range
of the squeezing parameter along with the success
probability for non-Gaussian operations providing positive
results.

\begin{widetext}
\begin{center}
\begin{table}[h!] 
\centering
\caption{\label{table1}
Advantageous range of squeezing parameter and success probability.}
\renewcommand{\arraystretch}{1.5}
\begin{tabular}{ |c |c |c |c|}
\hline
Operation & Squeezing Range (coh)& Squeezing Range (sqv)& Success Probability\\
\hline \hline
Asym 1-PS & $r \in [0.0,0.4]$ for $T=0.9$ & $r \in
[0.0,0.2]$ for $T=0.9$ & $P\sim 10^{-2}$ at $r=0.2$\\ \hline 
Sym 1-PS & $r \in [0.0,0.8]$ for $T=0.9$ & $r \in [0.0,1.0]$
for $T=0.9$  & $P\sim 10^{-2}$ at $r=0.6$ \\ \hline
Sym 3-PS & $r \in [0.0,0.8]$ for $T=0.9$ & $r \in [0.0,1.1]$
for $T=0.9$   & $P\sim  10^{-5}$ at $r=0.6$  \\ \hline
\hline
Sym 3-PA & $\emptyset$ for all values of $T$ &$r \in
[0.6,2.0]$ for $T=0.99$  & $P\sim 10^{-8}$ at $r=1$  \\
\hline
\hline
\end{tabular}
\end{table}
\end{center}
\end{widetext}
\section{Conclusion}
\label{sec:conc}
In this paper, we analytically derived the Wigner
characteristic function for the families of PATMSC and PSTMSC
states.  Using the expression for the Wigner characteristic
function, we computed the expression of the fidelity of
quantum teleportation
for input coherent states and squeezed vacuum states. 
For the PS case, our analysis showed that the symmetric PS
operations turn out to be more advantageous compared to
the  asymmetric PS operations which was useful only  in the
low squeezing range. PA operations were in general, less
useful than PS operations for quantum teleportation and it was only the
symmetric 3-PA which resulted in a modest improvement over
the TMSC state for an input squeezed vacuum. We were able
to recover the previous results available in this context 
as special cases from our more general analysis.

We considered an explicit physical model to carry out the PS
and PA operations. As it turns out, the scheme
succeeds only with a certain probability, leading to a
reduction of available resources.  The success probability
for Sym 1-PS is of the order $10^{-2}$, while for Sym 3-PS
and Sym 3-PA, the orders drop to $10^{-5}$ and $10^{-8}$,
respectively. Therefore, for a realistic comparison we must
take into account these success probabilities which further
reduces the reason for using these operations. In any case,
given the overall picture,  we conclude that the Sym 1-PS
operation is the most profitable non-Gaussian operation for
the purpose of carrying out quantum teleportation. We stress
that a more efficient way of carrying out PS and PA
operations is desirable if we want to use them for
enhancement of the efficiency of QIP protocols based on CV
systems.

We expect that the current work, where we have explicitly
computed the Wigner characteristic function for PSTMSC and
PATMSC states will pave the way to better analyse the 
application of these non-Gaussian states in various other QIP
protocols such as QKD, quantum metrology, and entanglement
swapping.
\section*{Acknowledgement} A and C.K.  acknowledge the
financial support from {\bf
DST/ICPS/QuST/Theme-1/2019/General} Project number {\sf
Q-68}.  \appendix \section{Continuous variable systems,
symplectic transformations and phase space
description}\label{intro}

We represent an $n$-mode continuous variable quantum system
via $n$ pairs of Hermitian quadrature operators
$\hat{q}_i,\hat{p}_i$ ($i=1\,,\dots, n$) which can be
grouped together in the form of a column vector
as~\cite{arvind1995,Braunstein,adesso-2007,weedbrook-rmp-2012,adesso-2014}
\begin{equation}\label{eq:columreal}
\hat{ \xi}  =(\hat{ \xi}_i)= (\hat{q_{1}},\,
\hat{p_{1}} \dots, \hat{q_{n}}, 
\, \hat{p_{n}})^{T}, \quad i = 1,2, \dots ,2n.
\end{equation}
The bosonic commutation relation  between them can be
expressed in a compact 
form as ($\hbar$=1)
\begin{equation}\label{eq:ccr}
[\hat{\xi}_i, \hat{\xi}_j] = i \Omega_{ij}, \quad
(i,j=1,2,...,2n),
\end{equation}
where $\Omega$ is the 2$n$ $\times$ 2$n$ matrix given by
\begin{equation}
\Omega = \bigoplus_{k=1}^{n}\omega =  \begin{pmatrix}
\omega & & \\
& \ddots& \\
& & \omega
\end{pmatrix}, \quad \omega = \begin{pmatrix}
0& 1\\
-1&0 
\end{pmatrix}.
\end{equation}
We can also represent an $n$-mode continuous variable
quantum system via $n$-pairs of annihilation  $\hat{a}_i$
and creation operators $\hat{a}^{\dagger}_i$
$(i=1,2,...,n)$. The field annihilation and creation
operators $\hat{a}_i\, \text{and}\, {\hat{a}_i}^{\dagger}$
($i=1,2,\, \dots\, ,n$) can be expressed in terms of
quadrature operators as 
\begin{equation}\label{realtocom}
\hat{a}_i=   \frac{1}{\sqrt{2}}(\hat{q}_i+i\hat{p}_i),
\quad  \hat{a}^{\dagger}_i=
\frac{1}{\sqrt{2}}(\hat{q}_i-i\hat{p}_i).
\end{equation}

\par
\noindent{\bf Displacement operator \,:} Displacement
operator acting on the $i^{\text{th}}$ mode is defined as
\begin{equation}\label{dis}
\hat{D}_i(a_i,b_i) = e^{i(b_i\hat{q}_i-a_i
\hat{p}_i)},
\end{equation}
where $a_i$ and $b_i$ is the amount of displacement along
$\hat{q}$ and $\hat{p}$-quadrature of the $i^{\text{th}}$
mode. Coherent state are generated by the action of the
displacement operator on the vacuum state:

\begin{equation}\label{coherent}
\vert a_i, b_i\rangle_i = \hat{D}_i(a_i,b_i)|0\rangle_i.
\end{equation}

Symplectic transformations are linear homogeneous
transformations, which are characterized by real $2n \times
2n$ matrices $S$ and transform the quadrature operators as
$\hat{\xi}_i \rightarrow \hat{\xi}_i^{\prime} =
S_{ij}\hat{\xi}_{j}$. The $S$ matrices obey the the
canonical commutation relations~(\ref{eq:ccr}) leading to
the condition $S\Omega S^T = \Omega$. These real
transformations constitute a non-compact group in $2n$
dimensions known as symplectic group denoted by
$Sp(2n,\,\mathcal{R})$. For each element $S \in
Sp(2n,\,\mathcal{R})$, there exists an infinite dimensional
unitary representation $\mathcal{U}(S)$  acting on the
Hilbert space. Now we define three symplectic operations,
which are relevant to this
work~\cite{arvind1995,weedbrook-rmp-2012}.

\par
\noindent{\bf Single mode squeezing operation \,:}
The single mode  squeezing operation transforms the
quadrature operators $(\hat{q},\hat{p})^T$ according to the
symplectic matrix
\begin{equation}\label{intro:sq}
S(r) = \begin{pmatrix}
e^{-r} & 0 \\
0 & e^{r}
\end{pmatrix}.
\end{equation}
The  infinite dimensional unitary representation for the
single mode 
squeezing operation can be written as
\begin{equation}\label{sq_one_mode}
\mathcal{U}(S(r)) = 
\exp[r (a^2-\hat{a}^{{\dagger}^2}) /2].
\end{equation}

\par
\noindent{\bf Beam splitter operation \,:}
The two mode beam splitter operation is a symplectic transformation whose 
action on the quadrature operators 
$  \hat{\xi} = (\hat{q}_{i}, \,\hat{p}_{i},\, \hat{q}_{j},\,
\hat{p}_{j})^{T}$ of a two mode system 
is given by
\begin{equation}\label{beamsplitter}
B_{ij}(T) = \begin{pmatrix}
\sqrt{T} \,\mathbb{1}_2& \sqrt{1-T} \,\mathbb{1}_2 \\
-\sqrt{1-T} \,\mathbb{1}_2& \sqrt{T} \,\mathbb{1}_2
\end{pmatrix},
\end{equation}
where $\mathbb{1}_2$ is the $2 \times 2$ identity matrix. The infinite 
dimensional unitary representation corresponding to the beam splitter 
transformation is
\begin{equation}\label{beam}
\mathcal{U}  (B_{ij}(T)) = \exp[\arccos(\sqrt{T})
(\hat{a}_i^{\dagger} \hat{a}_j-\hat{a}_i \hat{a}_j^{\dagger})].
\end{equation}

\par
\noindent{\bf  Two mode squeezing operation \,:}
The two mode squeezing operation is also a  symplectic transformation 
whose action on the quadrature operators 
$(\hat{q}_{i}$, $\hat{p}_{i}$, $\hat{q}_{j}$, $\hat{p}_{j})^T$ is given by
\begin{equation}\label{eq:tms}
S_{ij}(r) = \begin{pmatrix}
\cosh r \,\mathbb{1}_2& \sinh r \,\mathbb{Z} \\
\sinh r \,\mathbb{Z}& \cosh r \,\mathbb{1}_2
\end{pmatrix},
\end{equation}
where $\mathbb{Z} = \text{diag}(1,\, -1)$. The corresponding infinite 
dimensional unitary operator acting on the Hilbert space is
\begin{equation}\label{twomodesq}
\mathcal{U}  (S_{ij}(r)) = \exp[r(\hat{a}_i^{\dagger} \hat{a}_j^{\dagger}-\hat{a}_i
\hat{a}_j)].
\end{equation}
While the beam splitter operation  acting on the Hilbert space through its 
infinite dimensional unitary representation  $\mathcal{U}  (B_{ij}(\theta))$  
conserve the total photon number, single- and two-mode squeezing operator 
acting on the Hilbert space through its  infinite dimensional unitary 
representation  $\mathcal{U}  (S_{ij}(r))$  do not conserve the total photon 
number. We note that all the infinite dimensional unitary transformations of 
the three aforementioned symplectic transformations are generated by quadratic 
Hamiltonians.

\subsection{Phase space description}
For our current work, it is convenient to work in phase space formalism, 
more specifically, Wigner characteristic function. For a density operator 
$\hat{\rho}$ of an $n$-mode quantum system, the corresponding Wigner 
characteristic function is given by
\begin{equation}\label{wigdef}
\chi(\Lambda) = \text{Tr}[\hat{\rho} \, \exp(-i \Lambda^T \Omega \hat{\xi})],
\end{equation}
where $\xi = (\hat{q_1}, \hat{p_1},\dots \hat{q_n}, \hat{p_n})^T$,  
$\Lambda = (\Lambda_1, \Lambda_2, \dots \Lambda_n)^T$ with  
$\Lambda_i = (\tau_i, \sigma_i)^T \in \mathcal{R}^2$.
The Wigner characteristic function for a single mode Fock state 
$|n\rangle$ can be evaluated using Eq.~(\ref{wigdef}) as
\begin{equation}\label{charfock}
\chi_{|n\rangle}(\tau,\sigma)=\exp  \left[- \frac{\tau^2}{4}-\frac{\sigma^2}{4} \right]\,L_{n}\left( \frac{\tau^2}{2}+\frac{\sigma^2}{2} \right),
\end{equation}
where $L_n(x)$ is the Laguerre polynomial.
It can further be expressed as 
\begin{equation}\label{charfock1}
\chi_{|n\rangle}(\tau,\sigma)=\exp  \left[- \frac{\tau^2}{4}-\frac{\sigma^2}{4} \right]\,\bm{\widehat{F}}e^{ 2 st +s(\tau+i\sigma)-t(\tau-i\sigma)},
\end{equation}
with
\begin{equation}
\bm{\widehat{F}} =  \frac{1}{2^n n!}  \frac{\partial^n}{\partial\,s^n} \frac{\partial^n}{\partial\,t^n} \{ \bullet \}_{s=t=0}.
\end{equation}

For an $n$ mode system, the first order moments are defined as
\begin{equation}
\bm{d} = \langle  \hat{\xi } \rangle =
\text{Tr}[\hat{\rho} \hat{\xi}],
\end{equation}
and the second order moments, which can be written in the form of a 
real symmetric $2n\times2n$ matrix, is defined as
\begin{equation}\label{eq:cov}
V = (V_{ij})=\frac{1}{2}\langle \{\Delta \hat{\xi}_i,\Delta
\hat{\xi}_j\} \rangle,
\end{equation}
where $\Delta \hat{\xi}_i = \hat{\xi}_i-\langle \hat{\xi}_i
\rangle$, and $\{\,, \, \}$ denotes anti-commutator.
This matrix is called covariance matrix.
The uncertainty principle puts the following condition on the 
covariance matrix:
\begin{equation}
V+\frac{i}{2}\Omega \geq 0.
\end{equation}

A state is called a Gaussian state if the corresponding
Wigner characteristic function is a  Gaussian.  Gaussian states are
completely determined by their first and second order
moments.
For a Gaussian state, the Wigner characteristic function
is given by~\cite{weedbrook-rmp-2012, olivares-2012}
\begin{equation}\label{wigc}
\chi(\Lambda) =\exp[-\frac{1}{2}\Lambda^T (\Omega V \Omega^T) \Lambda- i (\Omega \bm{d} )^T\Lambda],
\end{equation}
where $V$ is the covariance matrix of the state and
$\bm{d}$ represents the displacement of the
Gaussian state. Using the above expression, the Wigner characteristic 
function of a single mode coherent state with displacement 
$\bm{d}=(d_x,d_p)^T$ turns out to be
\begin{equation}\label{chi_coh}
\chi_\text{coh}(\Lambda)= \exp \left[-\frac{1}{4}(\tau ^2+\sigma ^2)-i (\tau  d_p-\sigma  d_x)\right].
\end{equation}
Similarly, the characteristic function of a single mode squeezed 
vacuum state obtained by the action of a single mode squeezing 
operator~(\ref{sq_one_mode}) on vacuum state evaluates to
\begin{equation}\label{chi_sqv}
\chi_\text{sqv}(\Lambda)=\exp \left[-\frac{1}{4}(\tau ^2 e^{2r}+\sigma ^2 e^{-2r}) \right].
\end{equation}

For a homogeneous symplectic transformation $S$  and   its infinite 
dimensional unitary representation $\mathcal{U}(S)$, the density 
operator transforms as 
$\rho \rightarrow \,\mathcal{U}(S) \rho
\,\mathcal{U}(S)^{\dagger}$.
The corresponding transformation of the displacement vector, 
covariance matrix  and Wigner characteristic function is given 
by~\cite{arvind1995,olivares-2012,weedbrook-rmp-2012}
\begin{equation}\label{transformation} 
\bm{d}\rightarrow S \bm{d},\quad V\rightarrow SVS^T,\quad  \text{and} \,\,\chi(\Lambda) \rightarrow \chi(S^{-1}\Lambda).
\end{equation}
We conclude this brief introduction of CV systems and its phase 
space formalism with a note that most of the concepts discussed 
are available at length in Refs.~\cite{arvind1995,weedbrook-rmp-2012,adesso-2014}.

\section{Matrices and coefficients appearing in the Wigner characteristic function, 
success probability and the fidelity of QT using PSTMSC and PATMSC state.}\label{appps}

\subsection{Wigner characteristic function and probability expression of the PSTMSC state}
Here we provide the values of $a_i$  and $b_i$, which appear in the Wigner characteristic 
function~(\ref{eq4}) and probability expression~(\ref{probps}) of the PSTMSC state. 
The values for coefficients $a_i$ and $b_i$ are given by
\begin{equation}\label{ai}
\begin{aligned}
a_0=&b_0^{-1}  \exp \left(d^2 \left(\frac{t_1^2 +t_2^2}{b_0} -2\right) \right), \\
a_1=&-b_0^{-1}(\alpha ^2 r_1^2 t_2^2),\\
a_2=&b_2-\frac{r_1 t_2 \alpha}{b_0}(\alpha  t_1 t_2 (\tau _1+i \sigma _1)+i\beta  \left(i \tau _2+\sigma _2\right)),\\
a_3=&b_3+\frac{r_1 t_2 \alpha}{b_0}(\alpha  t_1 t_2 (\tau _1-i \sigma _1)-\beta  \left(\tau _2+i \sigma _2\right)),\\
a_4=&-b_0^{-1}(\alpha ^2 r_2^2 t_1^2),\\
a_5=&b_5-\frac{r_2 t_1 \alpha}{b_0}(\alpha  t_1 t_2 (\tau _2+i \sigma _2)+i\beta  \left(i \tau _1+\sigma _1\right)),\\
a_6=&b_6+\frac{r_2 t_1 \alpha}{b_0}(\alpha  t_1 t_2 (\tau _2-i  \sigma _2)-\beta  \left(\tau _1+i \sigma _1\right)),\\
a_7=&b_0^{-1}(\alpha  \beta  r_1 r_2),\\
\end{aligned}
\end{equation}
where,
\begin{equation}\label{ai2}
\begin{aligned}
b_0=&1+\alpha ^2 \left(1-t_1^2 t_2^2\right),\\ 
b_2=&\frac{r_1 d}{b_0}(i+1)\left(i \alpha  t_2^2-\beta\right),\\
b_3=&\frac{r_1 d}{b_0}(i+1)\left(i \beta-\alpha  t_2^2 \right),\\
b_5=&\frac{r_2 d}{b_0}(i+1) \left(i \alpha  t_1^2-\beta\right),\\
b_6=&\frac{r_2 d}{b_0}(i+1) \left(i \beta-\alpha  t_1^2 \right).\\
\end{aligned}
\end{equation}
Here  $t_i=\sqrt{T_i}$ and $r_i=\sqrt{1-T_i}$ ($i=1,2$). 
Further $\alpha=\sinh \, r$ and $\beta=\cosh \, r$. The matrix $M_1$ is given by
\begin{equation}\label{ai3}
M_1=\frac{-1}{4 b_0}\left(
\begin{array}{cccc}
c_0 & 0 & - 2\alpha  \beta  t_1 t_2 & 0 \\
0 & c_0 & 0 &  2\alpha  \beta  t_1 t_2 \\
-2 \alpha  \beta  t_1 t_2 & 0 & c_0 & 0 \\
0 &  2\alpha  \beta  t_1 t_2 & 0 & c_0\\
\end{array}
\right),
\end{equation}
where $c_0=1+\alpha^2(1+t_1^2t_2^2)$. The explicit form of matrix $M_2$ can be written as
\begin{equation}\label{ai33}
M_2=\frac{d}{i b_0}\left(
\begin{array}{c}
t_1 \left(\beta -\alpha  t_2^2\right) \\
-t_1 \left(\beta +\alpha  t_2^2\right) \\
t_2 \left(\beta -\alpha  t_1^2\right) \\
-t_2 \left(\beta +\alpha  t_1^2\right) \\
\end{array}
\right).
\end{equation}
\subsection*{Expressing Eq.~(\ref{eqchar}) in terms of two variable 
Hermite polynomials }
We first provide two identities concerning two variable Hermite 
polynomials and its differentiation:
\begin{equation}\label{hmn1}
\begin{aligned}
H_{m,n}(x,y)&=\sum_{i=0}^{\text{min}(m,n)}\frac{(-1)^i \,m!\,n!\,x^{m-i}y^{n-i}}{i!\,(m-i)!\,(n-i)!}\\
&=\frac{\partial^{m}}{\partial\,s^{m}} \frac{\partial^{n}}{\partial\,t^{n}}\exp(-st+sx+ty)\bigg|_{s=t=0}.\\
\end{aligned}
\end{equation}
and
\begin{equation}\label{hmn3}
\frac{\partial^{i}}{\partial\,x^{i}} \frac{\partial^{j}}{\partial\,y^{j}} H_{m,n}(x,y) = \frac{m!\,n!}{(m-i)!\,(n-j)!}H_{m-i,n-j}(x,y),
\end{equation}
Now we consider the part of Eq.~(\ref{eqchar}) that depends only 
on $u_i$ and $v_i$ ($i=1,2$) and arrange them as follows:
\begin{equation}\label{hmn2}
\begin{aligned}
&\bm{\widehat{F}_1} e^{a_7 \left(u_1 u_2+v_1 v_2\right)} e^{-a_1 u_1 v_1+a_2 u_1+a_3 v_1}e^{-a_4 u_2 v_2+a_5 u_2+a_6 v_2} \\
=&\bm{\widehat{F}_1} \sum_{i=0}^{\infty} \frac{(a_7 u_1 u_2)^i}{i!} \sum_{j=0}^{\infty} \frac{(a_7 v_1 v_2)^j}{j!}  e^{-a_1 u_1 v_1+a_2 u_1+a_3 v_1}\\
&\qquad \qquad \qquad \qquad \qquad \qquad \qquad \times e^{-a_4 u_2 v_2+a_5 u_2+a_6 v_2}  \\
=&\bm{\widehat{F}_1} \sum_{i=0}^{\infty} \frac{(a_7)^i}{i!} \sum_{j=0}^{\infty} \frac{(a_7)^j}{j!} \partial_{a_2}^{i} \partial_{a_5}^{i} \partial_{a_3}^{j} \partial_{a_6}^{j} e^{-a_1 u_1 v_1+a_2 u_1+a_3 v_1}\\
&\qquad \qquad \qquad \qquad \qquad \qquad \qquad \times e^{-a_4 u_2 v_2+a_5 u_2+a_6 v_2}  \\
=&\sum_{i,j=0}^{\infty}  \frac{(a_7)^{i+j}}{i!\,j!} \partial_{a_2}^{i} \partial_{a_5}^{i} \partial_{a_3}^{j} \partial_{a_6}^{j} a_1^{n_1} H_{n_1,n_1}\left[\frac{a_2}{\sqrt{a_1}},\frac{a_3}{\sqrt{a_1}}\right]\\
&\qquad \qquad \quad \qquad \qquad \qquad \times  a_4^{n_2} H_{n_2,n_2}\left[\frac{a_5}{\sqrt{a_4}},\frac{a_6}{\sqrt{a_4}}\right].\\
\end{aligned}
\end{equation}
Now we use Eq.(\ref{hmn2}) to obtain the final form (\ref{eq4}):
\begin{equation}\label{hmn4}
\begin{aligned}
&\sum_{i,j=0}^{\text{min}(n_1,n_2)}  \frac{(a_7)^{i+j}}{i!\,j!} a_1^{n_1}  \frac{P^{n_1}_i P^{n_1}_j}{\sqrt{a_1}^{i+j}} H_{n_1-i,n_1-j}\left[\frac{a_2}{\sqrt{a_1}},\frac{a_3}{\sqrt{a_1}}\right]\\
&\qquad \qquad \quad \,\,\, \times  a_4^{n_2} \frac{P^{n_2}_i P^{n_2}_j}{\sqrt{a_4}^{i+j}} H_{n_2-i,n_2-j}\left[\frac{a_5}{\sqrt{a_4}},\frac{a_6}{\sqrt{a_4}}\right].\\
\end{aligned}
\end{equation}

\subsection{Wigner characteristic function and probability expression of the PATMSC state}
Now we furnish the values of the coefficients $c_i$  and $d_i$   arising in 
the Wigner characteristic function~(\ref{unPA}) and probability 
expression~(\ref{probpa}) of the PATMSC state. The values for coefficients 
$c_i$ and $d_i$ are given as follows:
\begin{equation}\label{ai4}
\begin{aligned}
c_1=&- b_0^{-1} (\beta ^2 r_1^2),\\
c_2=&d_2-\frac{r_1\beta}{b_0} (\alpha  t_1 t_2 (\tau _2-i  \sigma _2)-\beta  \left(\tau _1+i \sigma _1\right))\\
c_3=&d_3+\frac{r_1\beta}{b_0} (\alpha  t_1 t_2 (\tau _2+i \sigma _2)+i\beta  \left(i \tau _1+\sigma _1\right))\\
c_4=&- b_0^{-1} (\beta ^2 r_2^2)\\
c_5=&d_5-\frac{r_2\beta}{b_0} (\alpha  t_1 t_2 (\tau _1-i \sigma _1)-\beta  \left(\tau _2+i \sigma _2\right))\\
c_6=&d_6+\frac{r_2\beta}{b_0} (\alpha  t_1 t_2 (\tau _1+i \sigma _1)+i\beta  \left(i \tau _2+\sigma _2\right))\\
c_7=& b_0^{-1} (\alpha  \beta  r_1 r_2 t_1 t_2)\\
\end{aligned}
\end{equation}
where, 
\begin{equation}\label{ai5}
\begin{aligned}
d_2=&-t_1 b_2, & \quad d_5=&-t_2 b_5,\\
d_3=&-t_1 b_3, & \quad d_6=&-t_2 b_6.\\
\end{aligned}
\end{equation}
\subsection{Fidelity for input coherent state using PSTMSC and PATMSC state}
The coefficients $e_i$ appearing in the fidelity of QT of input 
coherent state using PSTMSC state~(\ref{fidps}) are
\begin{equation}\label{ai6}
\begin{aligned}
e_0=&(b_0+d_0)^{-1} \exp \left(d^2\left(\frac{t_1^2+t_2^2}{b_0+d_0}-2\right)\right),\\
e_1=&-(b_0+d_0)^{-1}(\alpha ^2 r_1^2 t_2^2),\\
e_2=&\frac{r_1 d}{b_0+d_0} (i+1) \left(\alpha t_2 ( t_1 +i  t_2)-2 \beta\right),\\
e_3=&\frac{r_1 d}{b_0+d_0} (i+1) \left(2 i \beta -\alpha  t_2 (i  t_1 + t_2)\right),\\
e_4=&-(b_0+d_0)^{-1}(\alpha ^2 r_2^2 t_1^2),\\
e_5=&\frac{r_2 d}{b_0+d_0} (i+1) \left(\alpha  t_1 (i t_1+ t_2) -2 \beta\right),\\
e_6=&\frac{r_2 d}{b_0+d_0} (i+1) \left(2 i \beta -\alpha  t_1 (t_1+i  t_2) \right),\\
e_7=&(b_0+d_0)^{-1} \alpha  r_1 r_2 \left(2 \beta-\alpha  t_1 t_2 \right),\\
\end{aligned}
\end{equation}
where $d_0=(\beta-\alpha t_1 t_2)^2$. Further, the coefficients $f_i$ arising in 
the fidelity of QT of input coherent state using PATMSC state~(\ref{fidpa}) 
are given by
\begin{equation}\label{ai7}
\begin{aligned}
f_1=&-(b_0+d_0)^{-1}(\beta ^2 r_1^2),\\
f_2=&\frac{r_1 d \beta}{b_0+d_0} (i+1) \left(t_1-i t_2\right),\\
f_3=&\frac{r_1 d\beta}{b_0+d_0} (i+1) \left(t_2-i t_1\right),\\
f_4=&-(b_0+d_0)^{-1}(\beta ^2 r_2^2),\\
f_5=&\frac{r_2 d\beta}{b_0+d_0} (i+1) \left(t_2-i t_1\right),\\
f_6=&\frac{r_2 d\beta}{b_0+d_0} (i+1) \left(t_1-i t_2\right),\\
f_7=&(b_0+d_0)^{-1} \left(\beta ^2 r_1 r_2 \right).\\
\end{aligned}
\end{equation}
\subsection{Fidelity for input squeezed vacuum state using PSTMSC and PATMSC state}
The expression for $m_0$ in the fidelity of QT of input squeezed vacuum 
state using PSTMSC state~(\ref{sqvfidps}) is
\begin{equation}\label{ai8}
m_0= d^2\left(\frac{\left(t_1^2+t_2^2\right) \delta+2 t_1 t_2 \gamma+b_0 \left(t_1^2+t_2^2\right) d_0^{-1}}{2(c_0+b_0 \delta)}-2\right),
\end{equation}
with $\gamma=\sinh (2\epsilon)$, and $\delta=\cosh (2\epsilon)$. 
The matrix $M_3$ is given by
\begin{equation}\label{ai9}
M_3=\frac{1}{2(c_0+b_0 \delta)} \left(
\begin{array}{cccc}
g_1 & g_2 & g_3 & g_4 \\
g_2 & g_1 & g_4 & g_3 \\
g_3 & g_4 & g_5 & g_6 \\
g_4 & g_3 & g_6 & g_5 \\
\end{array}
\right),
\end{equation}
where
\begin{equation}
\begin{aligned}
g_1=&-\alpha ^2 r_1^2 t_2^2 \gamma,\\
g_2=& \alpha ^2 r_1^2 t_2^2 \left(\frac{b_0}{d_0}+\delta\right),\\
g_3=&\alpha r_1 r_2 \left(\frac{c_0}{\sqrt{d_0}}+\beta+\delta\left(\frac{b_0}{\sqrt{d_0}}+\beta\right)\right),\\
g_4=&2 \alpha ^2 r_1 r_2 t_1 t_2 \gamma,\\
g_5=&-\alpha ^2 r_2^2 t_1^2 \gamma,\\
g_6=& \alpha ^2 r_2^2 t_1^2 \left(\frac{b_0}{d_0}+\delta\right).\\
\end{aligned}
\end{equation}
The matrix $M_4$ is given by
\begin{equation}\label{ai10}
M_4=\frac{d(i+1)}{2(c_0+b_0 \delta)} \left(
\begin{array}{c}
\alpha r_1 t_2 (t_1-i t_2) g_7 + g_8 \\
\alpha r_1 t_2 (t_2-i t_1) g_9 - i g_8\\
\alpha r_2 t_1 (t_2-i t_1) g_7 + g_{10}  \\
\alpha r_2 t_1 (t_1-i t_2) g_9 - i g_{10}  \\
\end{array}
\right),
\end{equation}
where
\begin{equation}
\begin{aligned}
g_7=&\frac{b_0}{d_0}+\delta-i\gamma,& \quad g_9=
&\frac{b_0}{d_0}+\delta+i\gamma,\\
g_8=&2 \beta r_1 (1+\delta),& \quad g_{10}=&2 \beta r_2 (1+\delta).\\
\end{aligned}
\end{equation}
The expression for $M_5$ in the fidelity of QT of input squeezed vacuum 
state using PATMSC state~(\ref{sqvfidpa}) is
\begin{equation}\label{ai11}
M_5=\frac{1}{2(c_0+b_0 \delta)} \left(
\begin{array}{cccc}
h_1 & h_2 & h_3 & h_4 \\
h_2 & h_1 & h_4 & h_3 \\
h_3 & h_4 & h_5 & h_6 \\
h_4 & h_3 & h_6 & h_5 \\
\end{array}
\right),
\end{equation}
where,
\begin{equation}
\begin{aligned}
h_1=&-\beta ^2 r_1^2 \gamma,\\
h_2=&\beta ^2 r_1^2 \left(\frac{b_0}{d_0}+\delta\right),\\
h_3=&\beta r_1 r_2 \left(\frac{c_0}{\sqrt{d_0}}-\alpha t_1 t_2+\delta \left(\frac{b_0}{\sqrt{d_0}}+\alpha t_1 t_2\right)\right)\\
h_4=&\beta ^2 r_1 r_2 \gamma,\\
h_5=&-\beta ^2 r_2^2 \gamma,\\
h_6=&\beta ^2 r_2^2 \left(\frac{b_0}{d_0}+\delta \right).\\.\\
\end{aligned}
\end{equation}
The matrix $M_6$ is given by
\begin{equation}\label{ai12}
M_6=\frac{d(i+1)}{2(c_0+b_0 \delta)}\left(
\begin{array}{c}
r_1(t_1 h_7 -i t_2 h_8 +\beta \gamma(t_2+i t_1))\\
r_1(t_2 h_8 -i t_1 h_7 -\beta \gamma(t_1+i t_2))\\
r_2(t_2 h_7 -i t_1 h_8 +\beta \gamma(t_1+i t_2))\\
r_2(t_1 h_8 -i t_2 h_7 -\beta \gamma(t_2+i t_1))\\
\end{array}
\right),
\end{equation}
where
\begin{equation}
\begin{aligned}
h_7=&\beta \left(\frac{b_0}{d_0}+\delta\right), \,\, \text{and}\\
h_8=&\alpha t_1 t_2 \left(\frac{b_0}{d_0}+\delta\right) + b_0\left(\frac{d_0}{b_0}+\delta\right)d_0^{-1/2}. \\
\end{aligned}
\end{equation} 

\begin{widetext}
\end{widetext}


\begin{thebibliography}{49}%
	\makeatletter
	\providecommand \@ifxundefined [1]{%
		\@ifx{#1\undefined}
	}%
	\providecommand \@ifnum [1]{%
		\ifnum #1\expandafter \@firstoftwo
		\else \expandafter \@secondoftwo
		\fi
	}%
	\providecommand \@ifx [1]{%
		\ifx #1\expandafter \@firstoftwo
		\else \expandafter \@secondoftwo
		\fi
	}%
	\providecommand \natexlab [1]{#1}%
	\providecommand \enquote  [1]{``#1''}%
	\providecommand \bibnamefont  [1]{#1}%
	\providecommand \bibfnamefont [1]{#1}%
	\providecommand \citenamefont [1]{#1}%
	\providecommand \href@noop [0]{\@secondoftwo}%
	\providecommand \href [0]{\begingroup \@sanitize@url \@href}%
	\providecommand \@href[1]{\@@startlink{#1}\@@href}%
	\providecommand \@@href[1]{\endgroup#1\@@endlink}%
	\providecommand \@sanitize@url [0]{\catcode `\\12\catcode `\$12\catcode
		`\&12\catcode `\#12\catcode `\^12\catcode `\_12\catcode `\%12\relax}%
	\providecommand \@@startlink[1]{}%
	\providecommand \@@endlink[0]{}%
	\providecommand \url  [0]{\begingroup\@sanitize@url \@url }%
	\providecommand \@url [1]{\endgroup\@href {#1}{\urlprefix }}%
	\providecommand \urlprefix  [0]{URL }%
	\providecommand \Eprint [0]{\href }%
	\providecommand \doibase [0]{https://doi.org/}%
	\providecommand \selectlanguage [0]{\@gobble}%
	\providecommand \bibinfo  [0]{\@secondoftwo}%
	\providecommand \bibfield  [0]{\@secondoftwo}%
	\providecommand \translation [1]{[#1]}%
	\providecommand \BibitemOpen [0]{}%
	\providecommand \bibitemStop [0]{}%
	\providecommand \bibitemNoStop [0]{.\EOS\space}%
	\providecommand \EOS [0]{\spacefactor3000\relax}%
	\providecommand \BibitemShut  [1]{\csname bibitem#1\endcsname}%
	\let\auto@bib@innerbib\@empty
	\bibitem [{\citenamefont {Braunstein}\ and\ \citenamefont
		{Kimble}(1998)}]{bk-1998}%
	\BibitemOpen
	\bibfield  {author} {\bibinfo {author} {\bibfnamefont {S.~L.}\ \bibnamefont
			{Braunstein}}\ and\ \bibinfo {author} {\bibfnamefont {H.~J.}\ \bibnamefont
			{Kimble}},\ }\bibfield  {title} {\bibinfo {title} {Teleportation of
			continuous quantum variables},\ }\href
	{https://doi.org/10.1103/PhysRevLett.80.869} {\bibfield  {journal} {\bibinfo
			{journal} {Phys. Rev. Lett.}\ }\textbf {\bibinfo {volume} {80}},\ \bibinfo
		{pages} {869} (\bibinfo {year} {1998})}\BibitemShut {NoStop}%
	\bibitem [{\citenamefont {van Loock}\ and\ \citenamefont
		{Braunstein}(1999)}]{swapping-1999}%
	\BibitemOpen
	\bibfield  {author} {\bibinfo {author} {\bibfnamefont {P.}~\bibnamefont {van
				Loock}}\ and\ \bibinfo {author} {\bibfnamefont {S.~L.}\ \bibnamefont
			{Braunstein}},\ }\bibfield  {title} {\bibinfo {title} {Unconditional
			teleportation of continuous-variable entanglement},\ }\href
	{https://doi.org/10.1103/PhysRevA.61.010302} {\bibfield  {journal} {\bibinfo
			{journal} {Phys. Rev. A}\ }\textbf {\bibinfo {volume} {61}},\ \bibinfo
		{pages} {010302} (\bibinfo {year} {1999})}\BibitemShut {NoStop}%
	\bibitem [{\citenamefont {Laudenbach}\ \emph {et~al.}(2018)\citenamefont
		{Laudenbach}, \citenamefont {Pacher}, \citenamefont {Fung}, \citenamefont
		{Poppe}, \citenamefont {Peev}, \citenamefont {Schrenk}, \citenamefont
		{Hentschel}, \citenamefont {Walther},\ and\ \citenamefont
		{Hübel}}]{Laudenbach-review}%
	\BibitemOpen
	\bibfield  {author} {\bibinfo {author} {\bibfnamefont {F.}~\bibnamefont
			{Laudenbach}}, \bibinfo {author} {\bibfnamefont {C.}~\bibnamefont {Pacher}},
		\bibinfo {author} {\bibfnamefont {C.-H.~F.}\ \bibnamefont {Fung}}, \bibinfo
		{author} {\bibfnamefont {A.}~\bibnamefont {Poppe}}, \bibinfo {author}
		{\bibfnamefont {M.}~\bibnamefont {Peev}}, \bibinfo {author} {\bibfnamefont
			{B.}~\bibnamefont {Schrenk}}, \bibinfo {author} {\bibfnamefont
			{M.}~\bibnamefont {Hentschel}}, \bibinfo {author} {\bibfnamefont
			{P.}~\bibnamefont {Walther}},\ and\ \bibinfo {author} {\bibfnamefont
			{H.}~\bibnamefont {Hübel}},\ }\bibfield  {title} {\bibinfo {title}
		{Continuous-variable quantum key distribution with gaussian modulation—the
			theory of practical implementations},\ }\href
	{https://doi.org/https://doi.org/10.1002/qute.201800011} {\bibfield
		{journal} {\bibinfo  {journal} {Advanced Quantum Technologies}\ }\textbf
		{\bibinfo {volume} {1}},\ \bibinfo {pages} {1800011} (\bibinfo {year}
		{2018})}\BibitemShut {NoStop}%
	\bibitem [{\citenamefont {Walschaers}(2021)}]{Walschaers-prx-2021}%
	\BibitemOpen
	\bibfield  {author} {\bibinfo {author} {\bibfnamefont {M.}~\bibnamefont
			{Walschaers}},\ }\bibfield  {title} {\bibinfo {title} {Non-gaussian quantum
			states and where to find them},\ }\href
	{https://doi.org/10.1103/PRXQuantum.2.030204} {\bibfield  {journal} {\bibinfo
			{journal} {PRX Quantum}\ }\textbf {\bibinfo {volume} {2}},\ \bibinfo {pages}
		{030204} (\bibinfo {year} {2021})}\BibitemShut {NoStop}%
	\bibitem [{\citenamefont {Agarwal}\ and\ \citenamefont
		{Tara}(1991)}]{Agarwal-pra-1991}%
	\BibitemOpen
	\bibfield  {author} {\bibinfo {author} {\bibfnamefont {G.~S.}\ \bibnamefont
			{Agarwal}}\ and\ \bibinfo {author} {\bibfnamefont {K.}~\bibnamefont {Tara}},\
	}\bibfield  {title} {\bibinfo {title} {Nonclassical properties of states
			generated by the excitations on a coherent state},\ }\href
	{https://doi.org/10.1103/PhysRevA.43.492} {\bibfield  {journal} {\bibinfo
			{journal} {Phys. Rev. A}\ }\textbf {\bibinfo {volume} {43}},\ \bibinfo
		{pages} {492} (\bibinfo {year} {1991})}\BibitemShut {NoStop}%
	\bibitem [{\citenamefont {Kitagawa}\ \emph
		{et~al.}(2006{\natexlab{a}})\citenamefont {Kitagawa}, \citenamefont
		{Takeoka}, \citenamefont {Sasaki},\ and\ \citenamefont
		{Chefles}}]{Kitagawa-pra-2006}%
	\BibitemOpen
	\bibfield  {author} {\bibinfo {author} {\bibfnamefont {A.}~\bibnamefont
			{Kitagawa}}, \bibinfo {author} {\bibfnamefont {M.}~\bibnamefont {Takeoka}},
		\bibinfo {author} {\bibfnamefont {M.}~\bibnamefont {Sasaki}},\ and\ \bibinfo
		{author} {\bibfnamefont {A.}~\bibnamefont {Chefles}},\ }\bibfield  {title}
	{\bibinfo {title} {Entanglement evaluation of non-gaussian states generated
			by photon subtraction from squeezed states},\ }\href
	{https://doi.org/10.1103/PhysRevA.73.042310} {\bibfield  {journal} {\bibinfo
			{journal} {Phys. Rev. A}\ }\textbf {\bibinfo {volume} {73}},\ \bibinfo
		{pages} {042310} (\bibinfo {year} {2006}{\natexlab{a}})}\BibitemShut
	{NoStop}%
	\bibitem [{\citenamefont {Ourjoumtsev}\ \emph {et~al.}(2007)\citenamefont
		{Ourjoumtsev}, \citenamefont {Dantan}, \citenamefont {Tualle-Brouri},\ and\
		\citenamefont {Grangier}}]{Ourjoumtsev-prl-2007}%
	\BibitemOpen
	\bibfield  {author} {\bibinfo {author} {\bibfnamefont {A.}~\bibnamefont
			{Ourjoumtsev}}, \bibinfo {author} {\bibfnamefont {A.}~\bibnamefont {Dantan}},
		\bibinfo {author} {\bibfnamefont {R.}~\bibnamefont {Tualle-Brouri}},\ and\
		\bibinfo {author} {\bibfnamefont {P.}~\bibnamefont {Grangier}},\ }\bibfield
	{title} {\bibinfo {title} {Increasing entanglement between gaussian states by
			coherent photon subtraction},\ }\href
	{https://doi.org/10.1103/PhysRevLett.98.030502} {\bibfield  {journal}
		{\bibinfo  {journal} {Phys. Rev. Lett.}\ }\textbf {\bibinfo {volume} {98}},\
		\bibinfo {pages} {030502} (\bibinfo {year} {2007})}\BibitemShut {NoStop}%
	\bibitem [{\citenamefont {Takahashi}\ \emph {et~al.}(2010)\citenamefont
		{Takahashi}, \citenamefont {Neergaard-Nielsen}, \citenamefont {Takeuchi},
		\citenamefont {Takeoka}, \citenamefont {Hayasaka}, \citenamefont {Furusawa},\
		and\ \citenamefont {Sasaki}}]{Takahashi-nature-2010}%
	\BibitemOpen
	\bibfield  {author} {\bibinfo {author} {\bibfnamefont {H.}~\bibnamefont
			{Takahashi}}, \bibinfo {author} {\bibfnamefont {J.~S.}\ \bibnamefont
			{Neergaard-Nielsen}}, \bibinfo {author} {\bibfnamefont {M.}~\bibnamefont
			{Takeuchi}}, \bibinfo {author} {\bibfnamefont {M.}~\bibnamefont {Takeoka}},
		\bibinfo {author} {\bibfnamefont {K.}~\bibnamefont {Hayasaka}}, \bibinfo
		{author} {\bibfnamefont {A.}~\bibnamefont {Furusawa}},\ and\ \bibinfo
		{author} {\bibfnamefont {M.}~\bibnamefont {Sasaki}},\ }\bibfield  {title}
	{\bibinfo {title} {Entanglement distillation from gaussian input states},\
	}\href {https://doi.org/10.1038/nphoton.2010.1} {\bibfield  {journal}
		{\bibinfo  {journal} {Nature Photonics}\ }\textbf {\bibinfo {volume} {4}},\
		\bibinfo {pages} {178} (\bibinfo {year} {2010})}\BibitemShut {NoStop}%
	\bibitem [{\citenamefont {Zhang}\ and\ \citenamefont {van
			Loock}(2010)}]{Zhang-pra-2010}%
	\BibitemOpen
	\bibfield  {author} {\bibinfo {author} {\bibfnamefont {S.~L.}\ \bibnamefont
			{Zhang}}\ and\ \bibinfo {author} {\bibfnamefont {P.}~\bibnamefont {van
				Loock}},\ }\bibfield  {title} {\bibinfo {title} {Distillation of mixed-state
			continuous-variable entanglement by photon subtraction},\ }\href
	{https://doi.org/10.1103/PhysRevA.82.062316} {\bibfield  {journal} {\bibinfo
			{journal} {Phys. Rev. A}\ }\textbf {\bibinfo {volume} {82}},\ \bibinfo
		{pages} {062316} (\bibinfo {year} {2010})}\BibitemShut {NoStop}%
	\bibitem [{\citenamefont {Navarrete-Benlloch}\ \emph
		{et~al.}(2012)\citenamefont {Navarrete-Benlloch}, \citenamefont
		{Garc\'{\i}a-Patr\'on}, \citenamefont {Shapiro},\ and\ \citenamefont
		{Cerf}}]{pspra2012}%
	\BibitemOpen
	\bibfield  {author} {\bibinfo {author} {\bibfnamefont {C.}~\bibnamefont
			{Navarrete-Benlloch}}, \bibinfo {author} {\bibfnamefont {R.}~\bibnamefont
			{Garc\'{\i}a-Patr\'on}}, \bibinfo {author} {\bibfnamefont {J.~H.}\
			\bibnamefont {Shapiro}},\ and\ \bibinfo {author} {\bibfnamefont {N.~J.}\
			\bibnamefont {Cerf}},\ }\bibfield  {title} {\bibinfo {title} {Enhancing
			quantum entanglement by photon addition and subtraction},\ }\href
	{https://doi.org/10.1103/PhysRevA.86.012328} {\bibfield  {journal} {\bibinfo
			{journal} {Phys. Rev. A}\ }\textbf {\bibinfo {volume} {86}},\ \bibinfo
		{pages} {012328} (\bibinfo {year} {2012})}\BibitemShut {NoStop}%
	\bibitem [{\citenamefont {Tan}\ \emph {et~al.}(2008)\citenamefont {Tan},
		\citenamefont {Erkmen}, \citenamefont {Giovannetti}, \citenamefont {Guha},
		\citenamefont {Lloyd}, \citenamefont {Maccone}, \citenamefont {Pirandola},\
		and\ \citenamefont {Shapiro}}]{ill2008}%
	\BibitemOpen
	\bibfield  {author} {\bibinfo {author} {\bibfnamefont {S.-H.}\ \bibnamefont
			{Tan}}, \bibinfo {author} {\bibfnamefont {B.~I.}\ \bibnamefont {Erkmen}},
		\bibinfo {author} {\bibfnamefont {V.}~\bibnamefont {Giovannetti}}, \bibinfo
		{author} {\bibfnamefont {S.}~\bibnamefont {Guha}}, \bibinfo {author}
		{\bibfnamefont {S.}~\bibnamefont {Lloyd}}, \bibinfo {author} {\bibfnamefont
			{L.}~\bibnamefont {Maccone}}, \bibinfo {author} {\bibfnamefont
			{S.}~\bibnamefont {Pirandola}},\ and\ \bibinfo {author} {\bibfnamefont
			{J.~H.}\ \bibnamefont {Shapiro}},\ }\bibfield  {title} {\bibinfo {title}
		{Quantum illumination with gaussian states},\ }\href
	{https://doi.org/10.1103/PhysRevLett.101.253601} {\bibfield  {journal}
		{\bibinfo  {journal} {Phys. Rev. Lett.}\ }\textbf {\bibinfo {volume} {101}},\
		\bibinfo {pages} {253601} (\bibinfo {year} {2008})}\BibitemShut {NoStop}%
	\bibitem [{\citenamefont {Lopaeva}\ \emph {et~al.}(2013)\citenamefont
		{Lopaeva}, \citenamefont {Ruo~Berchera}, \citenamefont {Degiovanni},
		\citenamefont {Olivares}, \citenamefont {Brida},\ and\ \citenamefont
		{Genovese}}]{ill2013}%
	\BibitemOpen
	\bibfield  {author} {\bibinfo {author} {\bibfnamefont {E.~D.}\ \bibnamefont
			{Lopaeva}}, \bibinfo {author} {\bibfnamefont {I.}~\bibnamefont
			{Ruo~Berchera}}, \bibinfo {author} {\bibfnamefont {I.~P.}\ \bibnamefont
			{Degiovanni}}, \bibinfo {author} {\bibfnamefont {S.}~\bibnamefont
			{Olivares}}, \bibinfo {author} {\bibfnamefont {G.}~\bibnamefont {Brida}},\
		and\ \bibinfo {author} {\bibfnamefont {M.}~\bibnamefont {Genovese}},\
	}\bibfield  {title} {\bibinfo {title} {Experimental realization of quantum
			illumination},\ }\href {https://doi.org/10.1103/PhysRevLett.110.153603}
	{\bibfield  {journal} {\bibinfo  {journal} {Phys. Rev. Lett.}\ }\textbf
		{\bibinfo {volume} {110}},\ \bibinfo {pages} {153603} (\bibinfo {year}
		{2013})}\BibitemShut {NoStop}%
	\bibitem [{\citenamefont {Zhang}\ \emph {et~al.}(2014)\citenamefont {Zhang},
		\citenamefont {Guo}, \citenamefont {Bao}, \citenamefont {Shi}, \citenamefont
		{Jin}, \citenamefont {Zou},\ and\ \citenamefont {Guo}}]{illumination14}%
	\BibitemOpen
	\bibfield  {author} {\bibinfo {author} {\bibfnamefont {S.}~\bibnamefont
			{Zhang}}, \bibinfo {author} {\bibfnamefont {J.}~\bibnamefont {Guo}}, \bibinfo
		{author} {\bibfnamefont {W.}~\bibnamefont {Bao}}, \bibinfo {author}
		{\bibfnamefont {J.}~\bibnamefont {Shi}}, \bibinfo {author} {\bibfnamefont
			{C.}~\bibnamefont {Jin}}, \bibinfo {author} {\bibfnamefont {X.}~\bibnamefont
			{Zou}},\ and\ \bibinfo {author} {\bibfnamefont {G.}~\bibnamefont {Guo}},\
	}\bibfield  {title} {\bibinfo {title} {Quantum illumination with
			photon-subtracted continuous-variable entanglement},\ }\href
	{https://doi.org/10.1103/PhysRevA.89.062309} {\bibfield  {journal} {\bibinfo
			{journal} {Phys. Rev. A}\ }\textbf {\bibinfo {volume} {89}},\ \bibinfo
		{pages} {062309} (\bibinfo {year} {2014})}\BibitemShut {NoStop}%
	\bibitem [{\citenamefont {Fan}\ and\ \citenamefont
		{Zubairy}(2018)}]{illumination18}%
	\BibitemOpen
	\bibfield  {author} {\bibinfo {author} {\bibfnamefont {L.}~\bibnamefont
			{Fan}}\ and\ \bibinfo {author} {\bibfnamefont {M.~S.}\ \bibnamefont
			{Zubairy}},\ }\bibfield  {title} {\bibinfo {title} {Quantum illumination
			using non-gaussian states generated by photon subtraction and photon
			addition},\ }\href {https://doi.org/10.1103/PhysRevA.98.012319} {\bibfield
		{journal} {\bibinfo  {journal} {Phys. Rev. A}\ }\textbf {\bibinfo {volume}
			{98}},\ \bibinfo {pages} {012319} (\bibinfo {year} {2018})}\BibitemShut
	{NoStop}%
	\bibitem [{\citenamefont {Gupta}\ \emph {et~al.}(2024)\citenamefont {Gupta},
		\citenamefont {Roy}, \citenamefont {Das},\ and\ \citenamefont
		{Sen(De)}}]{rivu}%
	\BibitemOpen
	\bibfield  {author} {\bibinfo {author} {\bibfnamefont {R.}~\bibnamefont
			{Gupta}}, \bibinfo {author} {\bibfnamefont {S.}~\bibnamefont {Roy}}, \bibinfo
		{author} {\bibfnamefont {T.}~\bibnamefont {Das}},\ and\ \bibinfo {author}
		{\bibfnamefont {A.}~\bibnamefont {Sen(De)}},\ }\bibfield  {title} {\bibinfo
		{title} {Quantum illumination with noisy probes: Conditional advantages of
			non-gaussianity},\ }\href
	{https://doi.org/https://doi.org/10.1016/j.physleta.2024.129446} {\bibfield
		{journal} {\bibinfo  {journal} {Physics Letters A}\ }\textbf {\bibinfo
			{volume} {505}},\ \bibinfo {pages} {129446} (\bibinfo {year}
		{2024})}\BibitemShut {NoStop}%
	\bibitem [{\citenamefont {Opatrn\'y}\ \emph {et~al.}(2000)\citenamefont
		{Opatrn\'y}, \citenamefont {Kurizki},\ and\ \citenamefont
		{Welsch}}]{tel2000}%
	\BibitemOpen
	\bibfield  {author} {\bibinfo {author} {\bibfnamefont {T.}~\bibnamefont
			{Opatrn\'y}}, \bibinfo {author} {\bibfnamefont {G.}~\bibnamefont {Kurizki}},\
		and\ \bibinfo {author} {\bibfnamefont {D.-G.}\ \bibnamefont {Welsch}},\
	}\bibfield  {title} {\bibinfo {title} {Improvement on teleportation of
			continuous variables by photon subtraction via conditional measurement},\
	}\href {https://doi.org/10.1103/PhysRevA.61.032302} {\bibfield  {journal}
		{\bibinfo  {journal} {Phys. Rev. A}\ }\textbf {\bibinfo {volume} {61}},\
		\bibinfo {pages} {032302} (\bibinfo {year} {2000})}\BibitemShut {NoStop}%
	\bibitem [{\citenamefont {Kitagawa}\ \emph
		{et~al.}(2006{\natexlab{b}})\citenamefont {Kitagawa}, \citenamefont
		{Takeoka}, \citenamefont {Sasaki},\ and\ \citenamefont
		{Chefles}}]{Akira-pra-2006}%
	\BibitemOpen
	\bibfield  {author} {\bibinfo {author} {\bibfnamefont {A.}~\bibnamefont
			{Kitagawa}}, \bibinfo {author} {\bibfnamefont {M.}~\bibnamefont {Takeoka}},
		\bibinfo {author} {\bibfnamefont {M.}~\bibnamefont {Sasaki}},\ and\ \bibinfo
		{author} {\bibfnamefont {A.}~\bibnamefont {Chefles}},\ }\bibfield  {title}
	{\bibinfo {title} {Entanglement evaluation of non-gaussian states generated
			by photon subtraction from squeezed states},\ }\href
	{https://doi.org/10.1103/PhysRevA.73.042310} {\bibfield  {journal} {\bibinfo
			{journal} {Phys. Rev. A}\ }\textbf {\bibinfo {volume} {73}},\ \bibinfo
		{pages} {042310} (\bibinfo {year} {2006}{\natexlab{b}})}\BibitemShut
	{NoStop}%
	\bibitem [{\citenamefont {Dell'Anno}\ \emph {et~al.}(2007)\citenamefont
		{Dell'Anno}, \citenamefont {De~Siena}, \citenamefont {Albano},\ and\
		\citenamefont {Illuminati}}]{Anno-2007}%
	\BibitemOpen
	\bibfield  {author} {\bibinfo {author} {\bibfnamefont {F.}~\bibnamefont
			{Dell'Anno}}, \bibinfo {author} {\bibfnamefont {S.}~\bibnamefont {De~Siena}},
		\bibinfo {author} {\bibfnamefont {L.}~\bibnamefont {Albano}},\ and\ \bibinfo
		{author} {\bibfnamefont {F.}~\bibnamefont {Illuminati}},\ }\bibfield  {title}
	{\bibinfo {title} {Continuous-variable quantum teleportation with
			non-gaussian resources},\ }\href {https://doi.org/10.1103/PhysRevA.76.022301}
	{\bibfield  {journal} {\bibinfo  {journal} {Phys. Rev. A}\ }\textbf {\bibinfo
			{volume} {76}},\ \bibinfo {pages} {022301} (\bibinfo {year}
		{2007})}\BibitemShut {NoStop}%
	\bibitem [{\citenamefont {Yang}\ and\ \citenamefont {Li}(2009)}]{tel2009}%
	\BibitemOpen
	\bibfield  {author} {\bibinfo {author} {\bibfnamefont {Y.}~\bibnamefont
			{Yang}}\ and\ \bibinfo {author} {\bibfnamefont {F.-L.}\ \bibnamefont {Li}},\
	}\bibfield  {title} {\bibinfo {title} {Entanglement properties of
			non-gaussian resources generated via photon subtraction and addition and
			continuous-variable quantum-teleportation improvement},\ }\href
	{https://doi.org/10.1103/PhysRevA.80.022315} {\bibfield  {journal} {\bibinfo
			{journal} {Phys. Rev. A}\ }\textbf {\bibinfo {volume} {80}},\ \bibinfo
		{pages} {022315} (\bibinfo {year} {2009})}\BibitemShut {NoStop}%
	\bibitem [{\citenamefont {Wang}\ \emph {et~al.}(2015)\citenamefont {Wang},
		\citenamefont {Hou}, \citenamefont {Chen},\ and\ \citenamefont
		{Xu}}]{wang2015}%
	\BibitemOpen
	\bibfield  {author} {\bibinfo {author} {\bibfnamefont {S.}~\bibnamefont
			{Wang}}, \bibinfo {author} {\bibfnamefont {L.-L.}\ \bibnamefont {Hou}},
		\bibinfo {author} {\bibfnamefont {X.-F.}\ \bibnamefont {Chen}},\ and\
		\bibinfo {author} {\bibfnamefont {X.-F.}\ \bibnamefont {Xu}},\ }\bibfield
	{title} {\bibinfo {title} {Continuous-variable quantum teleportation with
			non-gaussian entangled states generated via multiple-photon subtraction and
			addition},\ }\href {https://doi.org/10.1103/PhysRevA.91.063832} {\bibfield
		{journal} {\bibinfo  {journal} {Phys. Rev. A}\ }\textbf {\bibinfo {volume}
			{91}},\ \bibinfo {pages} {063832} (\bibinfo {year} {2015})}\BibitemShut
	{NoStop}%
	\bibitem [{\citenamefont {Xu}(2015)}]{catalysis15}%
	\BibitemOpen
	\bibfield  {author} {\bibinfo {author} {\bibfnamefont {X.-x.}\ \bibnamefont
			{Xu}},\ }\bibfield  {title} {\bibinfo {title} {Enhancing quantum entanglement
			and quantum teleportation for two-mode squeezed vacuum state by local
			quantum-optical catalysis},\ }\href
	{https://doi.org/10.1103/PhysRevA.92.012318} {\bibfield  {journal} {\bibinfo
			{journal} {Phys. Rev. A}\ }\textbf {\bibinfo {volume} {92}},\ \bibinfo
		{pages} {012318} (\bibinfo {year} {2015})}\BibitemShut {NoStop}%
	\bibitem [{\citenamefont {Hu}\ \emph {et~al.}(2017)\citenamefont {Hu},
		\citenamefont {Liao},\ and\ \citenamefont {Zubairy}}]{catalysis17}%
	\BibitemOpen
	\bibfield  {author} {\bibinfo {author} {\bibfnamefont {L.}~\bibnamefont
			{Hu}}, \bibinfo {author} {\bibfnamefont {Z.}~\bibnamefont {Liao}},\ and\
		\bibinfo {author} {\bibfnamefont {M.~S.}\ \bibnamefont {Zubairy}},\
	}\bibfield  {title} {\bibinfo {title} {Continuous-variable entanglement via
			multiphoton catalysis},\ }\href {https://doi.org/10.1103/PhysRevA.95.012310}
	{\bibfield  {journal} {\bibinfo  {journal} {Phys. Rev. A}\ }\textbf {\bibinfo
			{volume} {95}},\ \bibinfo {pages} {012310} (\bibinfo {year}
		{2017})}\BibitemShut {NoStop}%
	\bibitem [{\citenamefont {Patra}\ \emph {et~al.}(2022)\citenamefont {Patra},
		\citenamefont {Gupta}, \citenamefont {Roy},\ and\ \citenamefont
		{Sen(De)}}]{ayan}%
	\BibitemOpen
	\bibfield  {author} {\bibinfo {author} {\bibfnamefont {A.}~\bibnamefont
			{Patra}}, \bibinfo {author} {\bibfnamefont {R.}~\bibnamefont {Gupta}},
		\bibinfo {author} {\bibfnamefont {S.}~\bibnamefont {Roy}},\ and\ \bibinfo
		{author} {\bibfnamefont {A.}~\bibnamefont {Sen(De)}},\ }\bibfield  {title}
	{\bibinfo {title} {Significance of fidelity deviation in continuous-variable
			teleportation},\ }\href {https://doi.org/10.1103/PhysRevA.106.022433}
	{\bibfield  {journal} {\bibinfo  {journal} {Phys. Rev. A}\ }\textbf {\bibinfo
			{volume} {106}},\ \bibinfo {pages} {022433} (\bibinfo {year}
		{2022})}\BibitemShut {NoStop}%
	\bibitem [{\citenamefont {Kumar}\ and\ \citenamefont
		{Arora}(2023)}]{tele-arxiv}%
	\BibitemOpen
	\bibfield  {author} {\bibinfo {author} {\bibfnamefont {C.}~\bibnamefont
			{Kumar}}\ and\ \bibinfo {author} {\bibfnamefont {S.}~\bibnamefont {Arora}},\
	}\bibfield  {title} {\bibinfo {title} {Success probability and performance
			optimization in non-gaussian continuous-variable quantum teleportation},\
	}\href {https://doi.org/10.1103/PhysRevA.107.012418} {\bibfield  {journal}
		{\bibinfo  {journal} {Phys. Rev. A}\ }\textbf {\bibinfo {volume} {107}},\
		\bibinfo {pages} {012418} (\bibinfo {year} {2023})}\BibitemShut {NoStop}%
	\bibitem [{\citenamefont {Kumar}\ \emph {et~al.}(2024)\citenamefont {Kumar},
		\citenamefont {Sharma},\ and\ \citenamefont {Arora}}]{better}%
	\BibitemOpen
	\bibfield  {author} {\bibinfo {author} {\bibfnamefont {C.}~\bibnamefont
			{Kumar}}, \bibinfo {author} {\bibfnamefont {M.}~\bibnamefont {Sharma}},\ and\
		\bibinfo {author} {\bibfnamefont {S.}~\bibnamefont {Arora}},\ }\bibfield
	{title} {\bibinfo {title} {Continuous variable quantum teleportation in a
			dissipative environment: Comparison of non-gaussian operations before and
			after noisy channel},\ }\href
	{https://doi.org/https://doi.org/10.1002/qute.202300344} {\bibfield
		{journal} {\bibinfo  {journal} {Advanced Quantum Technologies}\ }\textbf
		{\bibinfo {volume} {7}},\ \bibinfo {pages} {2300344} (\bibinfo {year}
		{2024})}\BibitemShut {NoStop}%
	\bibitem [{\citenamefont {Birrittella}\ \emph {et~al.}(2012)\citenamefont
		{Birrittella}, \citenamefont {Mimih},\ and\ \citenamefont
		{Gerry}}]{gerryc-pra-2012}%
	\BibitemOpen
	\bibfield  {author} {\bibinfo {author} {\bibfnamefont {R.}~\bibnamefont
			{Birrittella}}, \bibinfo {author} {\bibfnamefont {J.}~\bibnamefont {Mimih}},\
		and\ \bibinfo {author} {\bibfnamefont {C.~C.}\ \bibnamefont {Gerry}},\
	}\bibfield  {title} {\bibinfo {title} {Multiphoton quantum interference at a
			beam splitter and the approach to heisenberg-limited interferometry},\ }\href
	{https://doi.org/10.1103/PhysRevA.86.063828} {\bibfield  {journal} {\bibinfo
			{journal} {Phys. Rev. A}\ }\textbf {\bibinfo {volume} {86}},\ \bibinfo
		{pages} {063828} (\bibinfo {year} {2012})}\BibitemShut {NoStop}%
	\bibitem [{\citenamefont {Carranza}\ and\ \citenamefont
		{Gerry}(2012)}]{josab-2012}%
	\BibitemOpen
	\bibfield  {author} {\bibinfo {author} {\bibfnamefont {R.}~\bibnamefont
			{Carranza}}\ and\ \bibinfo {author} {\bibfnamefont {C.~C.}\ \bibnamefont
			{Gerry}},\ }\bibfield  {title} {\bibinfo {title} {Photon-subtracted two-mode
			squeezed vacuum states and applications to quantum optical interferometry},\
	}\href {https://doi.org/10.1364/JOSAB.29.002581} {\bibfield  {journal}
		{\bibinfo  {journal} {J. Opt. Soc. Am. B}\ }\textbf {\bibinfo {volume}
			{29}},\ \bibinfo {pages} {2581} (\bibinfo {year} {2012})}\BibitemShut
	{NoStop}%
	\bibitem [{\citenamefont {Braun}\ \emph {et~al.}(2014)\citenamefont {Braun},
		\citenamefont {Jian}, \citenamefont {Pinel},\ and\ \citenamefont
		{Treps}}]{braun-pra-2014}%
	\BibitemOpen
	\bibfield  {author} {\bibinfo {author} {\bibfnamefont {D.}~\bibnamefont
			{Braun}}, \bibinfo {author} {\bibfnamefont {P.}~\bibnamefont {Jian}},
		\bibinfo {author} {\bibfnamefont {O.}~\bibnamefont {Pinel}},\ and\ \bibinfo
		{author} {\bibfnamefont {N.}~\bibnamefont {Treps}},\ }\bibfield  {title}
	{\bibinfo {title} {Precision measurements with photon-subtracted or
			photon-added gaussian states},\ }\href
	{https://doi.org/10.1103/PhysRevA.90.013821} {\bibfield  {journal} {\bibinfo
			{journal} {Phys. Rev. A}\ }\textbf {\bibinfo {volume} {90}},\ \bibinfo
		{pages} {013821} (\bibinfo {year} {2014})}\BibitemShut {NoStop}%
	\bibitem [{\citenamefont {Ouyang}\ \emph {et~al.}(2016)\citenamefont {Ouyang},
		\citenamefont {Wang},\ and\ \citenamefont {Zhang}}]{josab-2016}%
	\BibitemOpen
	\bibfield  {author} {\bibinfo {author} {\bibfnamefont {Y.}~\bibnamefont
			{Ouyang}}, \bibinfo {author} {\bibfnamefont {S.}~\bibnamefont {Wang}},\ and\
		\bibinfo {author} {\bibfnamefont {L.}~\bibnamefont {Zhang}},\ }\bibfield
	{title} {\bibinfo {title} {Quantum optical interferometry via the
			photon-added two-mode squeezed vacuum states},\ }\href
	{https://doi.org/10.1364/JOSAB.33.001373} {\bibfield  {journal} {\bibinfo
			{journal} {J. Opt. Soc. Am. B}\ }\textbf {\bibinfo {volume} {33}},\ \bibinfo
		{pages} {1373} (\bibinfo {year} {2016})}\BibitemShut {NoStop}%
	\bibitem [{\citenamefont {Zhang}\ \emph {et~al.}(2021)\citenamefont {Zhang},
		\citenamefont {Ye}, \citenamefont {Wei}, \citenamefont {Xia}, \citenamefont
		{Chang}, \citenamefont {Liao},\ and\ \citenamefont
		{Hu}}]{pra-catalysis-2021}%
	\BibitemOpen
	\bibfield  {author} {\bibinfo {author} {\bibfnamefont {H.}~\bibnamefont
			{Zhang}}, \bibinfo {author} {\bibfnamefont {W.}~\bibnamefont {Ye}}, \bibinfo
		{author} {\bibfnamefont {C.}~\bibnamefont {Wei}}, \bibinfo {author}
		{\bibfnamefont {Y.}~\bibnamefont {Xia}}, \bibinfo {author} {\bibfnamefont
			{S.}~\bibnamefont {Chang}}, \bibinfo {author} {\bibfnamefont
			{Z.}~\bibnamefont {Liao}},\ and\ \bibinfo {author} {\bibfnamefont
			{L.}~\bibnamefont {Hu}},\ }\bibfield  {title} {\bibinfo {title} {Improved
			phase sensitivity in a quantum optical interferometer based on multiphoton
			catalytic two-mode squeezed vacuum states},\ }\href
	{https://doi.org/10.1103/PhysRevA.103.013705} {\bibfield  {journal} {\bibinfo
			{journal} {Phys. Rev. A}\ }\textbf {\bibinfo {volume} {103}},\ \bibinfo
		{pages} {013705} (\bibinfo {year} {2021})}\BibitemShut {NoStop}%
	\bibitem [{\citenamefont {Kumar}\ \emph {et~al.}(2022)\citenamefont {Kumar},
		\citenamefont {Rishabh},\ and\ \citenamefont {Arora}}]{crs-ngtmsv-met}%
	\BibitemOpen
	\bibfield  {author} {\bibinfo {author} {\bibfnamefont {C.}~\bibnamefont
			{Kumar}}, \bibinfo {author} {\bibnamefont {Rishabh}},\ and\ \bibinfo {author}
		{\bibfnamefont {S.}~\bibnamefont {Arora}},\ }\bibfield  {title} {\bibinfo
		{title} {Realistic non-gaussian-operation scheme in parity-detection-based
			mach-zehnder quantum interferometry},\ }\href
	{https://doi.org/10.1103/PhysRevA.105.052437} {\bibfield  {journal} {\bibinfo
			{journal} {Phys. Rev. A}\ }\textbf {\bibinfo {volume} {105}},\ \bibinfo
		{pages} {052437} (\bibinfo {year} {2022})}\BibitemShut {NoStop}%
	\bibitem [{\citenamefont {Kumar}\ \emph
		{et~al.}(2023{\natexlab{a}})\citenamefont {Kumar}, \citenamefont {Rishabh},
		\citenamefont {Sharma},\ and\ \citenamefont {Arora}}]{ngsvs}%
	\BibitemOpen
	\bibfield  {author} {\bibinfo {author} {\bibfnamefont {C.}~\bibnamefont
			{Kumar}}, \bibinfo {author} {\bibnamefont {Rishabh}}, \bibinfo {author}
		{\bibfnamefont {M.}~\bibnamefont {Sharma}},\ and\ \bibinfo {author}
		{\bibfnamefont {S.}~\bibnamefont {Arora}},\ }\bibfield  {title} {\bibinfo
		{title} {Parity-detection-based mach-zehnder interferometry with coherent and
			non-gaussian squeezed vacuum states as inputs},\ }\href
	{https://doi.org/10.1103/PhysRevA.108.012605} {\bibfield  {journal} {\bibinfo
			{journal} {Phys. Rev. A}\ }\textbf {\bibinfo {volume} {108}},\ \bibinfo
		{pages} {012605} (\bibinfo {year} {2023}{\natexlab{a}})}\BibitemShut
	{NoStop}%
	\bibitem [{\citenamefont {Kumar}\ \emph
		{et~al.}(2023{\natexlab{b}})\citenamefont {Kumar}, \citenamefont {Rishabh},\
		and\ \citenamefont {Arora}}]{metro-thermal}%
	\BibitemOpen
	\bibfield  {author} {\bibinfo {author} {\bibfnamefont {C.}~\bibnamefont
			{Kumar}}, \bibinfo {author} {\bibnamefont {Rishabh}},\ and\ \bibinfo {author}
		{\bibfnamefont {S.}~\bibnamefont {Arora}},\ }\bibfield  {title} {\bibinfo
		{title} {Enhanced phase estimation in parity-detection-based mach–zehnder
			interferometer using non-gaussian two-mode squeezed thermal input state},\
	}\href {https://doi.org/https://doi.org/10.1002/andp.202300117} {\bibfield
		{journal} {\bibinfo  {journal} {Annalen der Physik}\ }\textbf {\bibinfo
			{volume} {535}},\ \bibinfo {pages} {2300117} (\bibinfo {year}
		{2023}{\natexlab{b}})}\BibitemShut {NoStop}%
	\bibitem [{\citenamefont {Nha}\ and\ \citenamefont
		{Carmichael}(2004)}]{loophole-prl-2004}%
	\BibitemOpen
	\bibfield  {author} {\bibinfo {author} {\bibfnamefont {H.}~\bibnamefont
			{Nha}}\ and\ \bibinfo {author} {\bibfnamefont {H.~J.}\ \bibnamefont
			{Carmichael}},\ }\bibfield  {title} {\bibinfo {title} {Proposed test of
			quantum nonlocality for continuous variables},\ }\href
	{https://doi.org/10.1103/PhysRevLett.93.020401} {\bibfield  {journal}
		{\bibinfo  {journal} {Phys. Rev. Lett.}\ }\textbf {\bibinfo {volume} {93}},\
		\bibinfo {pages} {020401} (\bibinfo {year} {2004})}\BibitemShut {NoStop}%
	\bibitem [{\citenamefont {Garc\'{\i}a-Patr\'on}\ \emph
		{et~al.}(2004)\citenamefont {Garc\'{\i}a-Patr\'on}, \citenamefont
		{Fiur\'a\ifmmode~\check{s}\else \v{s}\fi{}ek}, \citenamefont {Cerf},
		\citenamefont {Wenger}, \citenamefont {Tualle-Brouri},\ and\ \citenamefont
		{Grangier}}]{Grangier-prl-2004}%
	\BibitemOpen
	\bibfield  {author} {\bibinfo {author} {\bibfnamefont {R.}~\bibnamefont
			{Garc\'{\i}a-Patr\'on}}, \bibinfo {author} {\bibfnamefont {J.}~\bibnamefont
			{Fiur\'a\ifmmode~\check{s}\else \v{s}\fi{}ek}}, \bibinfo {author}
		{\bibfnamefont {N.~J.}\ \bibnamefont {Cerf}}, \bibinfo {author}
		{\bibfnamefont {J.}~\bibnamefont {Wenger}}, \bibinfo {author} {\bibfnamefont
			{R.}~\bibnamefont {Tualle-Brouri}},\ and\ \bibinfo {author} {\bibfnamefont
			{P.}~\bibnamefont {Grangier}},\ }\bibfield  {title} {\bibinfo {title}
		{Proposal for a loophole-free bell test using homodyne detection},\ }\href
	{https://doi.org/10.1103/PhysRevLett.93.130409} {\bibfield  {journal}
		{\bibinfo  {journal} {Phys. Rev. Lett.}\ }\textbf {\bibinfo {volume} {93}},\
		\bibinfo {pages} {130409} (\bibinfo {year} {2004})}\BibitemShut {NoStop}%
	\bibitem [{\citenamefont {Bartlett}\ and\ \citenamefont
		{Sanders}(2002)}]{Bartlett-pra-2002}%
	\BibitemOpen
	\bibfield  {author} {\bibinfo {author} {\bibfnamefont {S.~D.}\ \bibnamefont
			{Bartlett}}\ and\ \bibinfo {author} {\bibfnamefont {B.~C.}\ \bibnamefont
			{Sanders}},\ }\bibfield  {title} {\bibinfo {title} {Universal
			continuous-variable quantum computation: Requirement of optical nonlinearity
			for photon counting},\ }\href {https://doi.org/10.1103/PhysRevA.65.042304}
	{\bibfield  {journal} {\bibinfo  {journal} {Phys. Rev. A}\ }\textbf {\bibinfo
			{volume} {65}},\ \bibinfo {pages} {042304} (\bibinfo {year}
		{2002})}\BibitemShut {NoStop}%
	\bibitem [{\citenamefont {Birrittella}\ \emph {et~al.}(2015)\citenamefont
		{Birrittella}, \citenamefont {Gura},\ and\ \citenamefont
		{Gerry}}]{tmsc-pra-2015}%
	\BibitemOpen
	\bibfield  {author} {\bibinfo {author} {\bibfnamefont {R.}~\bibnamefont
			{Birrittella}}, \bibinfo {author} {\bibfnamefont {A.}~\bibnamefont {Gura}},\
		and\ \bibinfo {author} {\bibfnamefont {C.~C.}\ \bibnamefont {Gerry}},\
	}\bibfield  {title} {\bibinfo {title} {Coherently stimulated parametric
			down-conversion, phase effects, and quantum-optical interferometry},\ }\href
	{https://doi.org/10.1103/PhysRevA.91.053801} {\bibfield  {journal} {\bibinfo
			{journal} {Phys. Rev. A}\ }\textbf {\bibinfo {volume} {91}},\ \bibinfo
		{pages} {053801} (\bibinfo {year} {2015})}\BibitemShut {NoStop}%
	\bibitem [{\citenamefont {Caves}\ \emph {et~al.}(1991)\citenamefont {Caves},
		\citenamefont {Zhu}, \citenamefont {Milburn},\ and\ \citenamefont
		{Schleich}}]{caves-pra-91}%
	\BibitemOpen
	\bibfield  {author} {\bibinfo {author} {\bibfnamefont {C.~M.}\ \bibnamefont
			{Caves}}, \bibinfo {author} {\bibfnamefont {C.}~\bibnamefont {Zhu}}, \bibinfo
		{author} {\bibfnamefont {G.~J.}\ \bibnamefont {Milburn}},\ and\ \bibinfo
		{author} {\bibfnamefont {W.}~\bibnamefont {Schleich}},\ }\bibfield  {title}
	{\bibinfo {title} {Photon statistics of two-mode squeezed states and
			interference in four-dimensional phase space},\ }\href
	{https://doi.org/10.1103/PhysRevA.43.3854} {\bibfield  {journal} {\bibinfo
			{journal} {Phys. Rev. A}\ }\textbf {\bibinfo {volume} {43}},\ \bibinfo
		{pages} {3854} (\bibinfo {year} {1991})}\BibitemShut {NoStop}%
	\bibitem [{\citenamefont {Selvadoray}\ \emph {et~al.}(1994)\citenamefont
		{Selvadoray}, \citenamefont {Kumar},\ and\ \citenamefont
		{Simon}}]{simon-pra-94}%
	\BibitemOpen
	\bibfield  {author} {\bibinfo {author} {\bibfnamefont {M.}~\bibnamefont
			{Selvadoray}}, \bibinfo {author} {\bibfnamefont {M.~S.}\ \bibnamefont
			{Kumar}},\ and\ \bibinfo {author} {\bibfnamefont {R.}~\bibnamefont {Simon}},\
	}\bibfield  {title} {\bibinfo {title} {Photon distribution in two-mode
			squeezed coherent states with complex displacement and squeeze parameters},\
	}\href {https://doi.org/10.1103/PhysRevA.49.4957} {\bibfield  {journal}
		{\bibinfo  {journal} {Phys. Rev. A}\ }\textbf {\bibinfo {volume} {49}},\
		\bibinfo {pages} {4957} (\bibinfo {year} {1994})}\BibitemShut {NoStop}%
	\bibitem [{\citenamefont {Marian}\ and\ \citenamefont
		{Marian}(2006)}]{Marian-pra-2006}%
	\BibitemOpen
	\bibfield  {author} {\bibinfo {author} {\bibfnamefont {P.}~\bibnamefont
			{Marian}}\ and\ \bibinfo {author} {\bibfnamefont {T.~A.}\ \bibnamefont
			{Marian}},\ }\bibfield  {title} {\bibinfo {title} {Continuous-variable
			teleportation in the characteristic-function description},\ }\href
	{https://doi.org/10.1103/PhysRevA.74.042306} {\bibfield  {journal} {\bibinfo
			{journal} {Phys. Rev. A}\ }\textbf {\bibinfo {volume} {74}},\ \bibinfo
		{pages} {042306} (\bibinfo {year} {2006})}\BibitemShut {NoStop}%
	\bibitem [{\citenamefont {Chizhov}\ \emph {et~al.}(2002)\citenamefont
		{Chizhov}, \citenamefont {Kn\"oll},\ and\ \citenamefont
		{Welsch}}]{Welsch-pra-2001}%
	\BibitemOpen
	\bibfield  {author} {\bibinfo {author} {\bibfnamefont {A.~V.}\ \bibnamefont
			{Chizhov}}, \bibinfo {author} {\bibfnamefont {L.}~\bibnamefont {Kn\"oll}},\
		and\ \bibinfo {author} {\bibfnamefont {D.-G.}\ \bibnamefont {Welsch}},\
	}\bibfield  {title} {\bibinfo {title} {Continuous-variable quantum
			teleportation through lossy channels},\ }\href
	{https://doi.org/10.1103/PhysRevA.65.022310} {\bibfield  {journal} {\bibinfo
			{journal} {Phys. Rev. A}\ }\textbf {\bibinfo {volume} {65}},\ \bibinfo
		{pages} {022310} (\bibinfo {year} {2002})}\BibitemShut {NoStop}%
	\bibitem [{\citenamefont {Braunstein}\ \emph {et~al.}(2000)\citenamefont
		{Braunstein}, \citenamefont {Fuchs},\ and\ \citenamefont
		{Kimble}}]{Braunstein-jmo-2000}%
	\BibitemOpen
	\bibfield  {author} {\bibinfo {author} {\bibfnamefont {S.~L.}\ \bibnamefont
			{Braunstein}}, \bibinfo {author} {\bibfnamefont {C.~A.}\ \bibnamefont
			{Fuchs}},\ and\ \bibinfo {author} {\bibfnamefont {H.~J.}\ \bibnamefont
			{Kimble}},\ }\bibfield  {title} {\bibinfo {title} {Criteria for
			continuous-variable quantum teleportation},\ }\href
	{https://doi.org/10.1080/09500340008244041} {\bibfield  {journal} {\bibinfo
			{journal} {Journal of Modern Optics}\ }\textbf {\bibinfo {volume} {47}},\
		\bibinfo {pages} {267} (\bibinfo {year} {2000})}\BibitemShut {NoStop}%
	\bibitem [{\citenamefont {Braunstein}\ \emph {et~al.}(2001)\citenamefont
		{Braunstein}, \citenamefont {Fuchs}, \citenamefont {Kimble},\ and\
		\citenamefont {van Loock}}]{Braunstein-pra-2001}%
	\BibitemOpen
	\bibfield  {author} {\bibinfo {author} {\bibfnamefont {S.~L.}\ \bibnamefont
			{Braunstein}}, \bibinfo {author} {\bibfnamefont {C.~A.}\ \bibnamefont
			{Fuchs}}, \bibinfo {author} {\bibfnamefont {H.~J.}\ \bibnamefont {Kimble}},\
		and\ \bibinfo {author} {\bibfnamefont {P.}~\bibnamefont {van Loock}},\
	}\bibfield  {title} {\bibinfo {title} {Quantum versus classical domains for
			teleportation with continuous variables},\ }\href
	{https://doi.org/10.1103/PhysRevA.64.022321} {\bibfield  {journal} {\bibinfo
			{journal} {Phys. Rev. A}\ }\textbf {\bibinfo {volume} {64}},\ \bibinfo
		{pages} {022321} (\bibinfo {year} {2001})}\BibitemShut {NoStop}%
	\bibitem [{\citenamefont {Arvind}\ \emph {et~al.}(1995)\citenamefont {Arvind},
		\citenamefont {Dutta}, \citenamefont {Mukunda},\ and\ \citenamefont
		{Simon}}]{arvind1995}%
	\BibitemOpen
	\bibfield  {author} {\bibinfo {author} {\bibnamefont {Arvind}}, \bibinfo
		{author} {\bibfnamefont {B.}~\bibnamefont {Dutta}}, \bibinfo {author}
		{\bibfnamefont {N.}~\bibnamefont {Mukunda}},\ and\ \bibinfo {author}
		{\bibfnamefont {R.}~\bibnamefont {Simon}},\ }\bibfield  {title} {\bibinfo
		{title} {The real symplectic groups in quantum mechanics and optics},\ }\href
	{https://doi.org/10.1007/BF02848172} {\bibfield  {journal} {\bibinfo
			{journal} {Pramana}\ }\textbf {\bibinfo {volume} {45}},\ \bibinfo {pages}
		{471} (\bibinfo {year} {1995})}\BibitemShut {NoStop}%
	\bibitem [{\citenamefont {Braunstein}\ and\ \citenamefont {van
			Loock}(2005)}]{Braunstein}%
	\BibitemOpen
	\bibfield  {author} {\bibinfo {author} {\bibfnamefont {S.~L.}\ \bibnamefont
			{Braunstein}}\ and\ \bibinfo {author} {\bibfnamefont {P.}~\bibnamefont {van
				Loock}},\ }\bibfield  {title} {\bibinfo {title} {Quantum information with
			continuous variables},\ }\href {https://doi.org/10.1103/RevModPhys.77.513}
	{\bibfield  {journal} {\bibinfo  {journal} {Rev. Mod. Phys.}\ }\textbf
		{\bibinfo {volume} {77}},\ \bibinfo {pages} {513} (\bibinfo {year}
		{2005})}\BibitemShut {NoStop}%
	\bibitem [{\citenamefont {Adesso}\ and\ \citenamefont
		{Illuminati}(2007)}]{adesso-2007}%
	\BibitemOpen
	\bibfield  {author} {\bibinfo {author} {\bibfnamefont {G.}~\bibnamefont
			{Adesso}}\ and\ \bibinfo {author} {\bibfnamefont {F.}~\bibnamefont
			{Illuminati}},\ }\bibfield  {title} {\bibinfo {title} {Entanglement in
			continuous-variable systems: recent advances and current perspectives},\
	}\href {https://doi.org/10.1088/1751-8113/40/28/S01} {\bibfield  {journal}
		{\bibinfo  {journal} {J. Phys. A}\ }\textbf {\bibinfo {volume} {40}},\
		\bibinfo {pages} {7821} (\bibinfo {year} {2007})}\BibitemShut {NoStop}%
	\bibitem [{\citenamefont {Weedbrook}\ \emph {et~al.}(2012)\citenamefont
		{Weedbrook}, \citenamefont {Pirandola}, \citenamefont {Garc\'{\i}a-Patr\'on},
		\citenamefont {Cerf}, \citenamefont {Ralph}, \citenamefont {Shapiro},\ and\
		\citenamefont {Lloyd}}]{weedbrook-rmp-2012}%
	\BibitemOpen
	\bibfield  {author} {\bibinfo {author} {\bibfnamefont {C.}~\bibnamefont
			{Weedbrook}}, \bibinfo {author} {\bibfnamefont {S.}~\bibnamefont
			{Pirandola}}, \bibinfo {author} {\bibfnamefont {R.}~\bibnamefont
			{Garc\'{\i}a-Patr\'on}}, \bibinfo {author} {\bibfnamefont {N.~J.}\
			\bibnamefont {Cerf}}, \bibinfo {author} {\bibfnamefont {T.~C.}\ \bibnamefont
			{Ralph}}, \bibinfo {author} {\bibfnamefont {J.~H.}\ \bibnamefont {Shapiro}},\
		and\ \bibinfo {author} {\bibfnamefont {S.}~\bibnamefont {Lloyd}},\ }\bibfield
	{title} {\bibinfo {title} {Gaussian quantum information},\ }\href
	{https://doi.org/10.1103/RevModPhys.84.621} {\bibfield  {journal} {\bibinfo
			{journal} {Rev. Mod. Phys.}\ }\textbf {\bibinfo {volume} {84}},\ \bibinfo
		{pages} {621} (\bibinfo {year} {2012})}\BibitemShut {NoStop}%
	\bibitem [{\citenamefont {Adesso}\ \emph {et~al.}(2014)\citenamefont {Adesso},
		\citenamefont {Ragy},\ and\ \citenamefont {Lee}}]{adesso-2014}%
	\BibitemOpen
	\bibfield  {author} {\bibinfo {author} {\bibfnamefont {G.}~\bibnamefont
			{Adesso}}, \bibinfo {author} {\bibfnamefont {S.}~\bibnamefont {Ragy}},\ and\
		\bibinfo {author} {\bibfnamefont {A.~R.}\ \bibnamefont {Lee}},\ }\bibfield
	{title} {\bibinfo {title} {Continuous variable quantum information:
			{G}aussian states and beyond},\ }\href
	{https://doi.org/10.1142/S1230161214400010} {\bibfield  {journal} {\bibinfo
			{journal} {Open Syst. Inf. Dyn.}\ }\textbf {\bibinfo {volume} {21}},\
		\bibinfo {pages} {1440001, 47} (\bibinfo {year} {2014})}\BibitemShut
	{NoStop}%
	\bibitem [{\citenamefont {Olivares}(2012)}]{olivares-2012}%
	\BibitemOpen
	\bibfield  {author} {\bibinfo {author} {\bibfnamefont {S.}~\bibnamefont
			{Olivares}},\ }\bibfield  {title} {\bibinfo {title} {Quantum optics in the
			phase space},\ }\href {https://doi.org/10.1140/epjst/e2012-01532-4}
	{\bibfield  {journal} {\bibinfo  {journal} {The European Physical Journal
				Special Topics}\ }\textbf {\bibinfo {volume} {203}},\ \bibinfo {pages} {3}
		(\bibinfo {year} {2012})}\BibitemShut {NoStop}%
\end{thebibliography}
%

\end{document}